\documentclass[fleqn]{2017SCGE}
\usepackage{mwe,tikz,ulem}
\usepackage{enumitem}                
\setlist{                            
  listparindent=\parindent,          
  parsep=0pt,                        
  itemsep=0pt,                       
  topsep=0pt,                        
  leftmargin=3.5mm                   
}
\usepackage{graphicx}

\usepackage{color}
\begin{document}

\ensubject{subject}

\ArticleType{Article}
\SpecialTopic{SPECIAL TOPIC: }
\Year{2017}
\Month{January}
\Vol{60}
\No{1}
\DOI{10.1007/s11432-016-0037-0}
\ArtNo{000000}
\ReceiveDate{January 11, 2016}
\AcceptDate{April 6, 2016}

\title{Accretion in Strong Field Gravity with \textit{eXTP}}


\author[]{\parbox[t]{17.5cm}{
Alessandra De Rosa$^{1}$,
Phil Uttley$^{2}$,
Lijun Gou$^{3}$,
Yuan Liu$^{4}$,
Cosimo Bambi$^{5}$,
Didier Barret$^{6}$,
Tomaso Belloni$^{7}$,
Emanuele Berti$^{8}$,
Stefano Bianchi$^{9}$,
Ilaria Caiazzo$^{10}$
% ...
}}{}

\address[]{\parbox[t]{18cm}{
$^{1}$INAF -- Istituto di Astrofisica e Planetologie Spaziali, Via Fosso del Cavaliere, 00133 Rome, Italy
$^{2}$Anton Pannekoek Institute for Astronomy, University of Amsterdam, Science Park 904, 1098 XH, Amsterdam, the Netherlands
$^{3}$National Astronomical Observatories, Chinese Academy of Sciences, Beijing 100012, China
$^{4}$Institute of High Energy Physics, Chinese Academy of Science, PO Box 918-3, Beijing 100049, China
$^{5}$Center for Field Theory and Particle Physics and Department of Physics, Fudan University, 220 Handan Road, 200433 Shanghai, China
$^{6}$IRAP, Universit\'e de Toulouse, CNRS, UPS, CNES, Toulouse, France 
$^{7}$INAF -- Osservatorio Astronomico di Brera, via E. Bianchi 46, 23807 Merate, Italy
$^{8}$Department of Physics and Astronomy, The University of Mississippi, University, Mississippi 38677, USA
$^{9}$Dipartimento di Matematica e Fisica, Universit\'a degli studi Roma Tre, Via della Vasca Navale 84, 00146 Rome, Italy
$^{10}$University of British Columbia, Vancouver, BC, V6T 1Z1, Canada
% ...
}}{}

\AuthorMark{A. De Rosa}

\AuthorCitation{A. De Rosa, P. Uttley, L. Gou et al}


\abstract{In this paper we describe the potential of the \textit{enhanced X-ray Timing and Polarimetry 
(eXTP)} mission for studies related to accretion flows in the strong field gravity regime around both 
stellar-mass and supermassive black-holes. {\it eXTP} has the unique capability of using 
advanced ``spectral-timing-polarimetry'' techniques to analyze the rapid variations with three orthogonal 
diagnostics of the flow and its geometry, yielding unprecedented insight into the 
inner accreting regions, the effects of strong field gravity on the material within them and the 
powerful outflows which are driven by the accretion process.}
%

\maketitle


\begin{multicols}{2}
\section{Introduction}\label{sect:section1}

One of the major challenges of modern astrophysics is the study of matter close to the event horizon of black holes (BH). The motion of accreting plasma near super-massive black holes (SMBHs) hosted in Active Galactic Nuclei (AGN) and stellar-mass black holes in X-ray binaries (XRBs), provides a powerful diagnostic to study the very deep potential well generated by the central object.  In the widely accepted scenario, the infalling matter forms an accretion disk that may extend down to the innermost stable circular orbit (ISCO), in the vicinity of which the bulk of the X-ray radiation is emitted. X-ray timing, spectroscopic and polarimetric techniques for probing matter flows into the strong gravity regime have been developed and, the first two, applied to real data. 

X-ray measurements in the strong field gravity regime can be used to infer the two most fundamental black hole parameters: mass and spin.  The black hole spin plays a major role in modern astrophysics; for instance it may provide an important source of energy, sustain winds and jets in AGN and stellar mass black holes, power the inner engine of gamma ray bursts and help to explain the apparent radio-loud/radio-quiet `dichotomy' in AGN.  Understanding the distribution of spins is crucial for understanding black hole formation and growth, giving an insight into the earlier universe (for SMBH) or (for stellar mass black holes) the physics of super- or hyper-novae \cite{volonterietal2005}.
Furthermore, since the spacetime is dominated by a single black hole which is accreting a negligible proportion of its mass, observations of matter close to the black hole can be used to verify some of the key predictions of General Relativity (GR) in a very different - and complementary - setting to that probed using gravitational wave measurements of black hole mergers. 
Gravitational wave detectors \cite{sathyaprakashetal2009} detect compact-object inspiral and merger events, where spacetime is being shaken by closely orbiting masses and hence is dynamic. X-ray observations of accreting black holes will probe instead the stationary spacetime metrics.

Strong field gravity effects in neutron stars are an important topic of study also, but in this White Paper, we focus on the strong-field gravity regions around black holes for two reasons.  Firstly, these objects offer a more pristine environment to study the behaviour of the accreting gas, free from the effects of a strong magnetic field or solid surface.  Secondly, astrophysical black holes cover, uniquely, 8-9 orders of magnitude in object mass, allowing a unique test of the scale invariance of gravitational effects and a link between the behaviour of accretion flows in stellar mass systems, the state changes of which can be studied in only days or weeks, versus those around supermassive black holes where we see only a snapshot of the current accretion state, which likely only changes on time-scales much longer than a human lifetime.  Equivalently, the study of AGN can inform us about the most rapid individual variations around accreting black holes in much greater detail than can be probed in XRBs, where we detect far fewer photons per light-crossing time (gaining sensitivity in XRBs only because we can average over many more cycles of variability).  However, before continuing our focus on black holes, we stress that many of the techniques which we apply to BH XRBs here, will be equally applicable to studying accreting neutron stars.

\subsection{Strong field gravity diagnostics}
To date, the two most important direct diagnostics of matter behaviour in the strong-field gravity regime in XRBs and AGN are (1) relativistically broadened Fe lines \cite{fabianetal1989,ref2,ref17} and (2) relativistic time-scale variability, in particular, quasi-periodic oscillations, QPOs \cite{stellaetal1999,ref4,ref5,ref6,ref7,ref8,ref9}.  

\begin{figure}[H]
\centering
\includegraphics[scale=0.3, angle=-90]{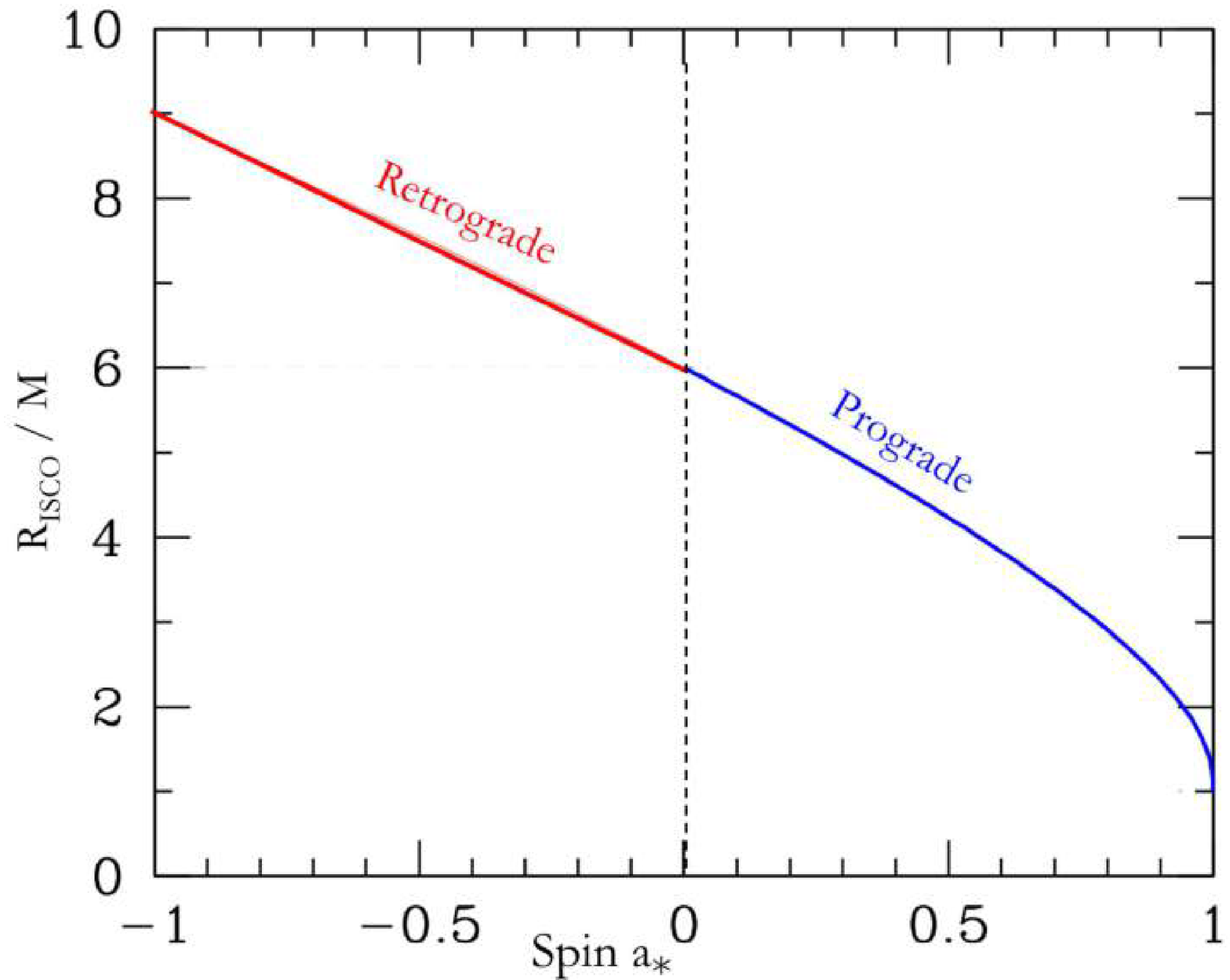}
\caption{The relation between the ISCO radius and the black hole spin.} 
\label{fig:radius_spin}
\end{figure}

The Fe K$\alpha$ emission line, along with the so-called `reflection continuum' (a broad bump starting at a few keV and peaking around 30 keV), is produced when the disk is externally illuminated, e.g. by the hot Comptonizing gas (often referred to as the `corona') which is responsible for the power-law X-ray emission component of AGN and Galactic black hole systems. An intrinsically narrow line emitted locally in the inner disk is hugely broadened and distorted by a combination of special relativistic  (Doppler effect and relativistic aberration) and GR (gravitational redshift, light bending) effects. By comparing the data to the line profile that is obtained by integrating over the line emitting disk region, it is possible to measure the accretion disk parameters like the innermost radius and the inclination angle   \cite{ref10,ref11}. If the inner disk radius corresponds to the ISCO, then since the ISCO radius depends on the black hole spin (see ~\cref{fig:radius_spin}), the latter can be inferred.
Very broad and asymmetric Fe K profiles around $\sim$ 5-7 keV have been observed in the X-ray spectra of bright AGN and XRBs \cite{ref12,ref13,ref14,ref15,ref16}.  

As a further check, the spin can also be determined in black hole XRBs by measuring the disk inner radius using model fitting of the thermal emission coming from the accretion disk itself \cite{ref28,ref18}, which is visible in the X-ray band for the high-temperature disks in these stellar mass systems. 

X-ray QPOs are routinely observed \cite{vanderKlis2006} in stellar mass black holes 
(and neutron stars) at low (0.1--10~Hz) and high (few hundred Hz) frequencies, within 
10-20\% of the relativistic precession, orbital and epicyclic frequencies in the inner 
disk, i.e., the fundamental frequencies of orbital motion in strong-field GR \cite{ref62,stellaetal1999}.  
The low and high-frequency QPOs have sometimes been seen together, in combinations of frequencies consistent 
with those expected from the multiple relativistic signals associated with a given radius in the accretion 
flow \cite{Mottaetal2014a,Mottaetal2014b}.  Thus, QPOs may correspond to oscillations excited 
in narrow ranges of radii and are potentially very strong probes of relativistic dynamics which 
also provide measurements of black hole spin, one of the main parameters which determines the frequencies.

\subsection{A complex astrophysical environment}
\label{ssect:complexenv}
In practice these spectral and timing signals of strong-field gravity are embedded in a more complex 
astrophysical environment, in some cases consisting of a hot, variable, geometrically thick and possibly 
optically thin inner accretion flow which is likely to occur close to the compact object and may correspond 
to the X-ray power-law emitting `corona' \cite{Doneetal2007}. The central region  is certainly more complex 
than a `standard' thin accretion disk, and possibly partly masked by accretion-powered outflows and 
(in the case of AGN), a complex gaseous environment.  The additional astrophysical effects pose challenges 
for application of standard spectroscopic and timing methods to study strong-field gravity effects.

\begin{figure*}[ht!]
\centering
\includegraphics[scale=0.5,angle=-90]{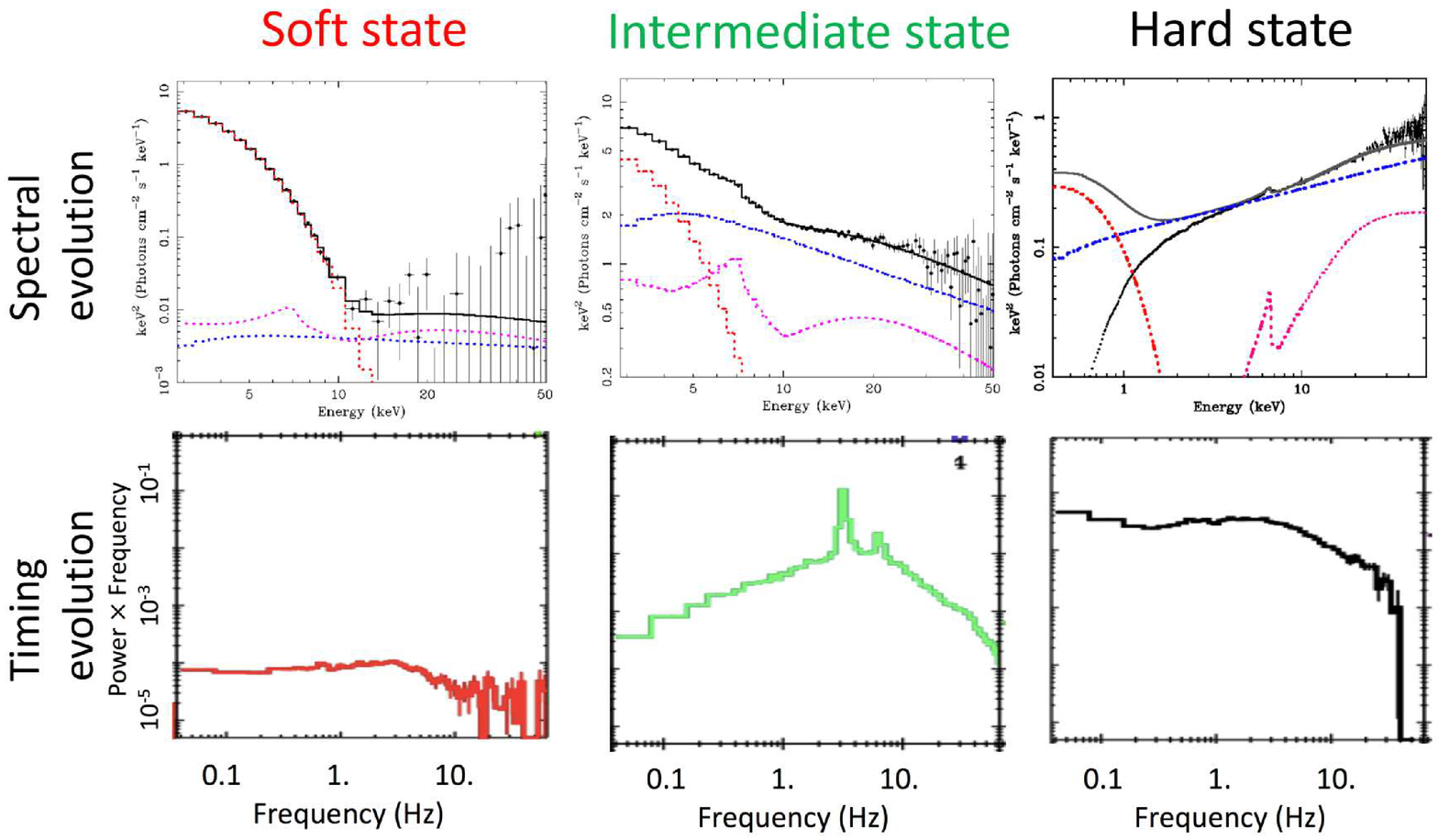}
\vspace{-0.5cm}
\caption{The distinct spectral and timing properties of the soft, intermediate and hard states (representative examples from the {\it Rossi X-ray Timing Explorer}, {\it RXTE} and {\it XMM-Newton} data).  The spectra (top row) show the disk blackbody (red), power-law (blue) and disk relativistic reflection (purple) components that best fit the data. (black points).  The hard state spectrum shows the components unabsorbed by Galactic absorption, revealing the disk blackbody at low energies. The states are characterised by a strongly changing ratio of disk blackbody to power-law emission, with the disk strongest in the soft state and weakest in the hard state.  These changes suggest a transfer of power between the disk and the corona, which may also correspond to truncation of the inner disk into an inner `hot flow', which might also produce low-frequency QPOs via precession.  The Fourier power-spectra (bottom row), obtained by {\it RXTE}, show that the hard and soft states are dominated by broadband noise variability, while the intermediate states tend to be dominated by strong low-frequency QPOs.} 
\label{fig:xrbstates}
\end{figure*}

Firstly, the spectral estimators of black hole spin described above rely on the assumption that the disk 
inner radius is located at the ISCO, but this situation is unlikely to apply in all observed cases.  Most 
notably, many BH XRBs are transients, showing strong outburst behaviour and strong spectral and timing evolution 
through different `states' during their outbursts (see~\cite{ref48,Doneetal2007} and  \cref{fig:xrbstates}), 
which suggests evolution of the inner accretion flow which produces most of the emission, and possibly corresponding 
changes in inner disk radius, as power is transferred between the disk and the corona.  As indicated in 
\cref{fig:xrbstates}, low-frequency QPOs are strongest in the intermediate states where the emission is not 
clearly dominated by either disk or corona.  High-frequency QPOs are seen in an even more restricted range 
of luminous intermediate states \cite{ref49,Motta2016}, but with current data it is not clear whether this 
restriction is a real physical effect or simply an observational bias due to them being stronger and also 
easier to detect in these very bright states.  Our interpretation of the QPOs is strongly limited unless 
we can understand whether and how they arise in distinct spectral states and what the actual changes in 
emission geometry are that correspond to the changes in state.

Furthermore, when modelling relativistically broadened Fe emission, it is imperative to have a good estimate of the underlying broad continuum shape \cite{ref17}, but AGN in particular often show spectra which are significantly modified by intervening absorption or distant reflection components.  For example, reflection in AGN may also arise from neutral/ionised Compton-thick gas, like the pc-scale torus envisaged by unification models \cite{antonucci93}. Absorption may be due to accretion disk winds, discovered and studied in AGN and XRBs for more than a decade \cite{ref17,ref79,ref90,ref91}.  These may take the form of AGN Ultra-Fast Outflows (UFOs) with velocities $\sim0.1c$ \cite{ref81}, observable at locations of sub-parsec scales from the central super-massive black hole, suggesting an identification with a disk wind or the base of a possible weak/broad jet. These components could provide a significant contribution to the observed AGN feedback between the central supermassive black hole and its host galaxy \cite{dimatteoetal05}.

Also, in AGN we may see both neutral absorption from intervening cold matter  (due to e.g. the Broad Line Regions, the absorbing torus or host galaxy dust lanes) and ionised absorption from intervening warm matter with different velocities. These latter absorbers are often outflowing with velocities of hundreds/thousands km~s$^{-1}$ (the standard `warm absorber').

X-ray polarization signatures are also expected from the regions close to the black hole from several physical components, but to date these signatures have been unexplored due to the lack of sensitive polarimetric capability on previous and current X-ray observatories.  Firstly, electron-scattering in the accretion disk should produce polarized disk thermal emission with a polarization fraction of up to a few per cent in the disk-dominated soft states of black hole XRBs.  Furthermore during the propagation of X-ray photons in the strong gravitational field the effects of relativistic beaming, gravitational lensing and frame dragging can lead to the  rotation of the integrated polarization vector \cite{connors&stark77,connorsetal80,matt93,dovciaketal08,ref72}. 
Another potential polarized signal from close to accreting black holes is the central corona (possibly corresponding to the `inner hot flow' of matter in the innermost disk). The geometry and emission mechanisms of the corona are still mysterious and X-ray polarization observations can provide a fundamental probe of the coronal geometry, which is helpful for understanding its physical origin.  For example, it is trivial to determine whether a corona is oriented in a plane above the disk or has a more central spherical geometry \cite{ref74}.
Nevertheless, we expect polarized signals from the disk and corona to be combined and 
this will make it difficult for stand-alone polarimeters to distinguish between the 
different components and make the kind of breakthroughs that are opened up with this 
important new capability.  

\subsection{The breakthrough capabilities of {\it eXTP}}
To tackle these challenges and fully unlock the potential of X-ray observations of black holes to study the behaviour of matter in strong field gravity, the {\it enhanced X-ray Timing and Polarimetry} mission ({\it eXTP}) has been proposed by a consortium led by the Institute of High-Energy Physics of the Chinese Academy of Sciences and envisaged for a launch in the mid 2020s. It
carries 4 instrument packages for the 0.5-–50 keV bandpass,
with the primary purpose to study conditions of extreme density \cite{wattsetal18},
gravity (this paper) and magnetism \cite{fengetal18} in and around compact objects
in the universe.  It will also be a powerful observatory for
a wider range of astrophysical phenomena since it combines
high throughput, good spectral and timing resolution, polarimetric
capability and wide sky coverage \cite{intzandetal18}.

The scientific payload of {\it eXTP} consists of:
the Spectroscopic Focusing Array (SFA), the Polarimetry Focusing
Array (PFA), the Large Area Detector (LAD) and the
Wide Field Monitor (WFM).  The {\it eXTP} instrumentation is discussed in detail in \cite{zhangetal18}, but here we give a brief overview along with the breakthroughs that will be provided by {\it eXTP}'s unique combination of instrument capabilities.

{\it eXTP}'s large-area and fast-timing and spectroscopy capability is provided across a broad X-ray energy range by the combination of the SFA, an array
of nine identical X-ray mirror and silicon drift detector combinations,
covering the  0.5–-10 keV energy range with a spectral resolution of better than 180 eV (full width at half maximum, FWHM) at 6 keV, and the LAD, a set of large-area collimated silicon drift detectors which cover the 2--50~keV range and have 260 eV resolution at 6 keV. The LAD and SFA together reach a total effective area of $\sim$ 4 m$^2$ at 6~keV and since both use silicon drift detectors, they are capable of sampling extremely high count rates at high time resolution (10~$\mu$s) with minimal deadtime.  The broad X-ray energy coverage and CCD-quality spectral resolution allows spectral fits to disentangle complex spectra, such as additional absorption and broad reflection continuum features in AGN and the combination of disk blackbody, coronal continuum and reflection in XRBs, thus enabling much better measurements of relativistic reflection features in AGN and XRBs (Sect.~\ref{ssect:reflection}) and the disk thermal emission in XRBs (Sect.~\ref{ssect:continuum_fitting}).  Measurements of the larger-scale absorber and reprocessor properties will themselves provide valuable information on the outflows of AGN and XRBs and the surrounding environment of AGN (Sect.~\ref{sect:astro}).

The large collecting area and fast timing capability also enable differential spectroscopy and spectral-timing techniques to be applied. Thanks to these techniques we can cleanly separate the spectral components from the innermost strong-gravity region,  variable on short time-scales, from those at larger scales which will vary much more slowly. 
For example, quasi-periodic variations can occur due to a precessing inner flow or orbiting inhomogeneities in the disk, and can be analyzed using Doppler tomography techniques, where red- and blueshifts are used to reconstruct the illumination pattern or loci of the inhomogeneities (Sect.~\ref{ssect:QPOtomo} and \ref{ssect:AGNtomo}). Black hole masses of AGN can also be inferred from variability in the Fe K-line (Sect.~\ref{ssect:AGNtomo}).   Reverberation (radiation ‘echoing’) of the variability of an incident hard continuum from the corona off the disk leads to light travel time lags between the different components, which manifest as distinct features in plots of lag vs. energy. These lags constrain the geometry on an absolute length scale (km), and in particular, diagnose the absolute size of the inner radius of the reflecting disk (Sect.~\ref{ssect:reverberation}), allowing us to constrain black hole masses, as well as changes in the inner geometry associated with different accretion states.

Furthermore, the high count rates obtained by the SFA and LAD allow detection and tracking of BH XRB high-frequency QPOs down to low rms amplitudes, allowing us to determine their origin and use them as an independent diagnostic of the black hole spin and a potential test of the dynamics of matter close to the black hole (Sect.~\ref{ssect:qpo}).  Increased sensitivity to weaker HF QPOs may also open up their detection in much greater numbers and in a wider range of accretion states than sampled previously.

Another independent diagnostic of strong field gravity is offered by X-ray polarimetric measurements carried out by the PFA, which consists of four identical X-ray telescopes that
are sensitive between 2 and 10 keV, have an angular resolution
better than 30'' and a total effective area of $\sim$ 500 cm$^2$ at
3 keV (including the detector effciency). The PFA features Gas Pixel Detectors (GPDs) to measure X-ray polarisation, reaching a minimum detectable polarization
(MDP) of 5\% in 100 ks for a source with mCrab-level flux
3$\times$10$^{-11}$ erg s$^{-1}$ cm$^{-2}$. The spectral resolution is 1.8 keV at 6 keV.  While it offers stand-alone capability to make X-ray polarimetric measurements with greater sensitivity than any previous instrument, the true strength of the PFA for studying strong field gravity lies in its capability to measure polarization signals simultaneously with other independent diagnostics of strong field gravity (Sect.~\ref{ssect:continuum_fitting}), as well as combine the polarization signal with the flux-variability signal from the large-area detectors.  This combination opens up the possibility of `spectral-timing-polarimetry', to analyze the rapid variations with three orthogonal constraints on the flow and its geometry, namely, (i) spectroscopy yielding velocities and redshifts, (ii) timing of orbiting patterns revealing orbital periods and GR precession in the accretion flow and (iii) polarimetry providing clues to the geometry and additional GR effects, yielding unprecedented insight into the inner flow.  In fact {\it eXTP} can use the combination of polarization signal and energy-resolved flux-variability signal from the large-area detectors to separate the different polarized components according to how they correlate differently with the X-ray spectral and flux variability produced in the innermost regions, allowing the data to be used in entirely new ways (Sect.~\ref{ssect:QPOpol}, \ref{ssect:xrb coronae}).

The science payload is completed by the WFM, consisting of 6 coded-mask cameras covering 3.7~sr of the sky at a sensitivity of 4 mCrab for an exposure time of 1 d in the 2 to 50 keV energy range, and for a typical sensitivity of 0.2 mCrab combining 1 yr of observations outside the Galactic plane. The instrument will feature an angular resolution of a few arcminutes and will be endowed with an energy resolution of about 300~eV. The baseline for the observatory response time to targets of opportunity within the 50\% part of the sky accessible to {\it eXTP} at any one time is 4–-8 hours. Dependent on  the outcome of mission studies, this may improve to 1–-3 hours.  The monitoring capability of the WFM will be essential to identify and follow up outbursts of new and known transient XRBs, as well as target specific states of transient and persistent XRBs and AGN for detailed follow-up with the pointed instruments, enabling the wide range of strong field gravity studies described in this White Paper.

In the remaining Sections we will present {\it eXTP}'s  capability to explore physical phenomena in the regions close to black holes.  We will discuss  the three independent diagnostics provided by {\it eXTP} data, in order to infer either the behaviour of matter in the strong-field gravity regime (Sect.~\ref{sect:SFG}) and the accretion physics and geometry in the inner region around black holes (Sect.~\ref{sect:inner_region}).  
Moreover, {\it eXTP} will provide new important information in the wider context of the astrophysics of accreting black holes (Sect.~\ref{sect:astro}) and, in particular, of the new gravitational wave astrophysics (Sect.~\ref{sect:sect_GW}).

\section{Matter in the strong-field gravity regime}\label{sect:SFG}
This section describes how {\it eXTP} will revolutionise the measurement of the core diagnostics of the behaviour of matter in strong-field gravity, using spectroscopic, polarimetric and timing measurements to provide independent estimates of black hole spin and the effects of strong field gravity.  The techniques employed are relatively `standard', either using time-averaged measurements of spectral and polarization signals, or measurements of QPO signals with Fourier power-spectral techniques applied to a broad continuum bandpass. Nevertheless  all these techniques benefit from the large collecting area and/or new instrumental capabilities of {\it eXTP} and provide a powerful suite of separate diagnostics that is unique to the mission. In Sect.~\ref{sect:inner_region} we will further show how we can combine these different types of measurement using state-of-the-art differential spectroscopy, spectral-timing and spectral-timing-polarimetry techniques, to gain even more insight into the central regions and the effects of strong field gravity.

\subsection{Relativistically broadened reflection}
\label{ssect:reflection}
\begin{figure*}[ht!]
\centering
\includegraphics[scale=0.6,angle=90]{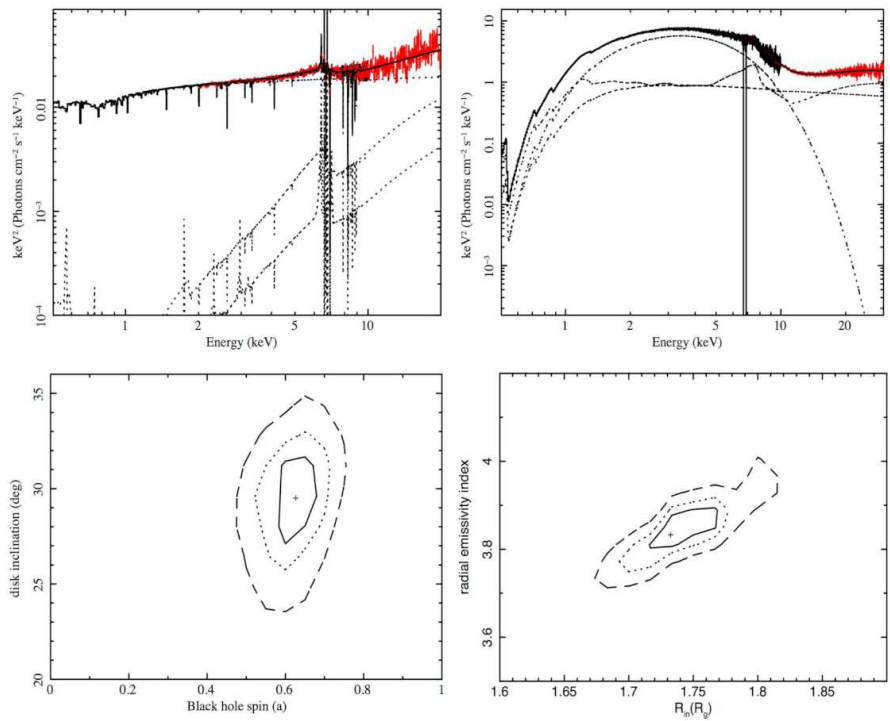}
\caption{Black hole spin measurements with {\it eXTP} (SFA in black and LAD in red). \textbf{Upper panels} \textit{Left}. {\it eXTP} spectrum as obtained by a 100 ks integration of a 2 mCrab, spin a=0.7 AGN  composed of: the continuum, three ionized absorber components, the cold reflection and narrow Fe line K$\alpha$, Fe K$\beta$ and Ni K$\alpha$, the ionized lines FeXXV and FeXXVI  and the blurred ionized reflection component. 
\textit{Right}. {\it eXTP} simulated spectrum as obtained in a 0.1 ks integration of a 0.5 Crab, spin $a=0.97$ XRB black hole. The simulated soft state spectrum includes a thermal disk component, hard Comptonized continuum, relativistic reflection features and highly photoionized absorption ($\log \xi$=3.6 for ionisation parameter $\xi$ in units of erg cm s$^{-1}$) due to an outflow producing narrow lines/edges in the Fe K region as observed in GRO~J1655$-$40  \cite{ref27}.
\textbf{Lower panels} Confidence levels for inclination vs. black hole spin for the AGN  (\textit{left}) and the radial emissivity index vs. inner radius (in units of the gravitational radius R$_g$=GM/c$^2$) for the XRB (\textit{right}), 1, 2 and 3 sigma as solid, dotted, and dashed lines respectively. }\label{fig:BH_spin}
\end{figure*}

The very broad Fe K$\alpha$ profiles at 6.4~keV often  observed in accreting black holes (both AGN and XRBs) and neutron stars are successfully modeled by X-ray reprocessing of a hard irradiating continuum by the accretion disk plasma in tight relativistic orbits around the compact object. In the case of black holes, the inner disk reflection models of increasing sophistication now include full Kerr-metric GR calculations of flow dynamics as well as photon trajectories and Doppler and gravitational redshifts, and an advanced treatment of the radiation processes \cite{ref22}. Models reproduce the broad Fe K line, the fluorescent emission features at lower energies and the Compton hump at energies above 10 keV. They allow us to study reflection of radiation described by different spectral slopes and radial distributions, from matter over a large range of ionizations, densities and chemical abundances, as a function of disk inclination and black hole spin \cite{Garcia2014,Garcia2016}.
In order to understand the emergent reflection spectrum, it is necessary also to understand the illumination pattern of the accretion disk, that is its emissivity profile, the reflected power per unit area as a function of location on the disk  \cite{ref23, ref24}. By comparing observed emissivity profiles to those computed theoretically for different locations and geometries of the source, it is possible to constrain the location and extent of the primary X-ray source (i.e. the emitting corona).
The Fe line profiles in X-ray spectra of the black hole systems thus provide a sensitive probe of the matter in the strong field region and estimates of black hole spin. Some current stellar as well as super-massive black hole spin estimates based on measuring time-averaged line profiles suggest near-maximal spins \cite{ref25}, but, as specificed above, there are complications related to spectral complexity \cite{milleretal09} and  pile-up effects (in XRBs) \cite{donediaz10}, so that significant discrepancies occur with respect to other techniques, e.g. disk continuum fitting (Sect. \ref{ssect:continuum_fitting})

The enormous S/N and good energy resolution available with {\it eXTP}, will allow us to measure average Fe line profiles with exceptional precision in both XRBs and AGN (as well as in neutron stars, as widely discussed in \cite{wattsetal18}), using state of the art reflection models to measure black hole spins. In \cref{fig:BH_spin} left panels, we show the  {\it eXTP} spectrum as obtained by a 100 ks integration of a bright 2 mCrab, spin a=0.7 AGN where the photoionized, relativistically broadened reflection component has been contaminated by contributions from three ionised absorber components as well as cold reflection and associated narrow Fe line K$\alpha$, Fe K$\beta$ and Ni K$\alpha$, as expected in typical type 1 AGN (see Sect. \ref{ssect:complexenv} and \ref{ssect:astro_agn}). Our simulations show that for AGN, the energy resolution of {\it eXTP} together with the large effective area and broadband energy coverage provided by the SFA and LAD combination, allow us to disentangle the spectral complexities in the Fe K region and measure the reflection continuum shape, to successfully extract the relativistic reflection parameters and recover the black hole spin with a precision of $\sim$10 per cent (\cref{fig:BH_spin} lower-left panel), despite the presence of the contaminating components.
In order to measur the black hole spin with high precision, we require a minimun S/N of 400 in the 2--10 keV band, this limit will allow  {\it eXTP} to carry out such detailed broad Fe line modeling on a large sample of AGN (more than 400 sources at different redshifts) with flux above $\sim$ 10$^{-12}$ erg cm$^{-2}$s$^{-1}$. 

In the case of stellar mass black holes {\it eXTP}, mainly thanks to the virtually pile-up free LAD data, will allow us to measure changes in the accretion flow structure close to black holes on unprecedentedly short time-scales.
In XRBs the disk inclination angle is usually obtained  from optical/IR observations so that the high S/N broadband data will allow to measure the disk emissivity profile thus putting strong constraints on the corona geometry.
In the simulation reported in ~\cref{fig:BH_spin}, right panels, we show the case of a 0.5 Crab object with maximal spin (such as the microquasar GRO~J1655$-$40 \cite{ref26,ref27}). An {\it eXTP} observation of Fe K emission in such a system will measure the inner radius of the disk and the index of the radial emissivity profile with a precision of (respectively) about 2 and 5 per cent in only 100 s. This result is obtained using the best current models including complex absorption (see ~\cref{fig:BH_spin} upper-right panel).  The unprecedentedly short timescale of such a measurement will allow us for the first time to observe the variability of the structure of the innermost region on a time-scale comparable to that of the fastest known changes in outflow components such as winds and jets, to open a completely new domain for the study of the inner regions around black holes and how the ejection properties are linked to the inner accretion flow (see also Sect.~\ref{ssect:astro_xrb}).

\subsection{The disk thermal emission in XRBs}
\label{ssect:continuum_fitting}
Besides producing the reflection features already discussed, the accretion disk also emits blackbody radiation produced by internal heating as well as external heating by the illuminating corona.  In AGN the disk thermal emission peaks in the UV band.  However, in stellar mass black holes accreting at the moderate to high rates seen in outbursting or persistent sources, the disk emission peaks in the X-ray band and provides an additional probe of the accretion flow close to the black hole.  In particular, in the disk-dominated soft state of black hole XRBs the disk is expected to reach the ISCO and hence a measurement of the disk inner radius could provide a direct measure of the black hole spin.  Since the disk emission is a superposition of blackbodies, the normalisation of the disk spectrum yields the emitting area and hence the radius: this is the so-called {\it continuum-fitting method} for estimating black hole spin \cite{ref28,ref29}.  
\begin{figure*}
\centering
\includegraphics[scale=0.4]{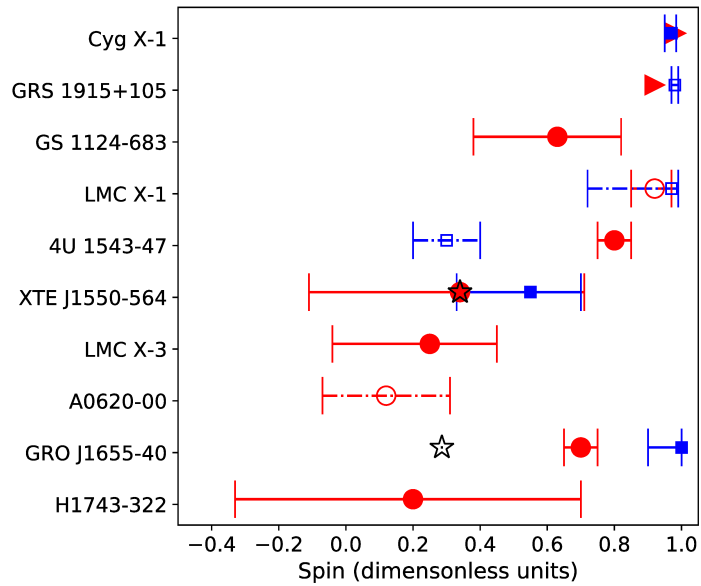}
\caption{Spin estimates from the literature for Galactic black holes (ordered by increasing black hole mass from bottom to top), with measurements made using the continuum fitting method (red circles or triangles).  Where they exist, estimates from relativistic Fe K line fitting (blue squares or triangles) and QPOs (black stars) are also given.  Triangles denote 3-$\sigma$ lower limits.  Solid symbols and error bars denote 90\% confidence intervals, while open symbols with dot-dashed error bars denote 68\% confidence intervals.  Errors on the QPO spin estimates are smaller than the data points and not shown.  Some significant discrepancies arise between the different techniques (e.g. for GRO~J1655-40), likely due to systematic errors in the current data and methods.  {\it eXTP}’s unique capabilities and combination of instruments and new diagnostics will substantially reduce these systematic errors.  References: {\it Continuum fitting:} Cyg X-1 \cite{ref33}, GRS 1915+105 \cite{ref35}, GS ~1124-683 \cite{ref39}, LMC X-1 \cite{ref37}, 4U 1543-47 \cite{ref41}, XTE~J1550-564 \cite{Steineretal2011}, LMC X-3 \cite{ref44}, A0620-00 \cite{ref46}, GRO~J1655-40 \cite{ref41}, H1743-322 \cite{ref45}.   {\it Fe~K fitting:} Cyg X-1 \cite{ref34}, GRS 1915+105 \cite{ref36}, LMC X-1 \cite{ref38}, 4U 1543-47 \cite{Miller2009}, XTE~J1550-564 \cite{Steineretal2011}, GRO~J1655-40 \cite{ref26}. {\it QPO method:} XTE~J1550-564 \cite{Mottaetal2014b} , GRO~J1655-40 \cite{Mottaetal2014a}.}
\label{fig:bhspins}
\end{figure*}

The fundamental assumptions underpinning the continuum-fitting method are already well-supported by observations and theory \cite{ref30,ref31}. 
To use the method to estimate the disk inner radius in gravitational units (and hence spin), it is essential to have good estimates of the source distance, the disk inclination and the black hole mass.  These system parameters are usually determined beforehand from optical/IR observations.  Currently, most distances are estimated from optical study of the companion star, which can be subject to significant uncertainties for sources in the galactic plane where extinction is large.  Measurements of Galactic structure and dynamics obtained from the {\it Gaia} mission should significantly reduce these uncertainties.  Also, in a recent breakthrough, much more accurate (to better than $\sim 10$~per~cent uncertainty) distances to XRBs have been obtained via Very Long Baseline Interferometry (VLBI) parallax measurements of their radio emission (e.g. \cite{ref35}).  By the mid-2020s, such measurements will be routine and made even more powerful by incorporating data from the Square Kilometre Array (SKA), so that errors on distance estimates should be reduced to $\sim 3$~per~cent (James Miller-Jones, private communication), leading to similar systematic uncertainties on the disk inner radius (and hence spin) from continuum fitting. 
\begin{figure*}[ht!]
\begin{minipage}{1\textwidth}
\centering
\includegraphics[scale=0.6]{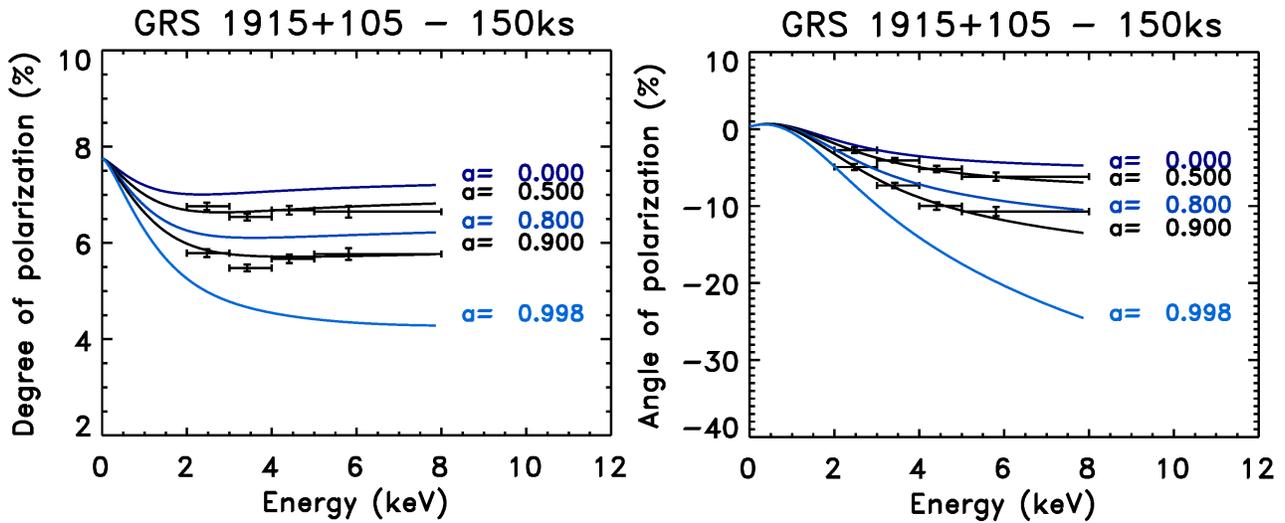}
\end{minipage}
\caption{Simulated {\it eXTP} PFA measurement of polarization degree (left panel) and angle (right panel) as a function of energy, expected from a 150~ks exposure of GRS~1915+105 in the soft state.  Blue and black curves show the expected dependencies of the polarization quantities on spin, while the black data points show the simulated data corresponding to the black curves.} \label{fig:pol_disk}
\end{figure*}

\cref{fig:bhspins} summarises the current state-of-the-art of BH XRB spin measurements using the continuum fitting method, in comparison with spin estimates from other techniques where available.  A wide range of spins have been measured, but in many cases errors are large and significant discrepancies arise with estimates of spin from iron line fitting, which may be a result of the systematic errors in both techniques.  However, in the mid-2020s {\it eXTP} will allow a big step forward in the accuracy of the use of disk thermal emission to map the innermost regions and measure black hole spin by combining spectral and polarimetric diagnostics.

The Wide Field Monitor (WFM) will first enable precise targeting of the soft states best suited for spin measurements, with minimal contamination by power-law emission.  The broad energy bandpass of the SFA and LAD will then allow an accurate determination of the disk spectral shape, with absorption features in the soft state accurately modelled (see Sect.~\ref{ssect:astro_xrb}) and the hard response of the LAD enabling a precise constraint on the hard power-law tail which can otherwise bias fits to the disk spectrum.

Due to the planar structure of the disk, its thermal emission is expected to be polarized due  to
Thomson scattering in the disk atmosphere.  Assuming a standard thin disk emission with a  Thomson scattering in the disk atmosphere as the origin of polarization, the black hole spin can then be obtained independently of disk continuum fitting, by measuring the rotation with energy of the polarization angle of the disk  emission. In fact, due to GR effects, the polarization plane of the disk radiation rotates while travelling along a geodesic. As a result, for a distant observer the plane of polarization is no longer parallel or perpendicular to the disk, as it would be in the Newtonian case. The rotation angle depends on the location of the emitting point in the disk, and it is larger the closer to the black hole the emitting point is. Since the temperature decreases with the disk radius, higher energy photons suffer a larger rotation. When the emission is integrated over the disk, a rotation of the polarization plane (together with variations in polarization degree) with energy is expected \cite{connors&stark77,connorsetal80,matt93,dovciaketal08,ref72}. The effect increases with the spin of the black hole, following the decrease with the spin of the ISCO radius; the spin can therefore be estimated via this effect (e.g. \cite{dovciaketal08}), as shown in Figure \ref{fig:pol_disk}.

\subsection{QPOs in X-ray binaries}
\label{ssect:qpo}
 \begin{figure*}[t]
\centering
\includegraphics[width=0.35\textwidth,angle=-90]{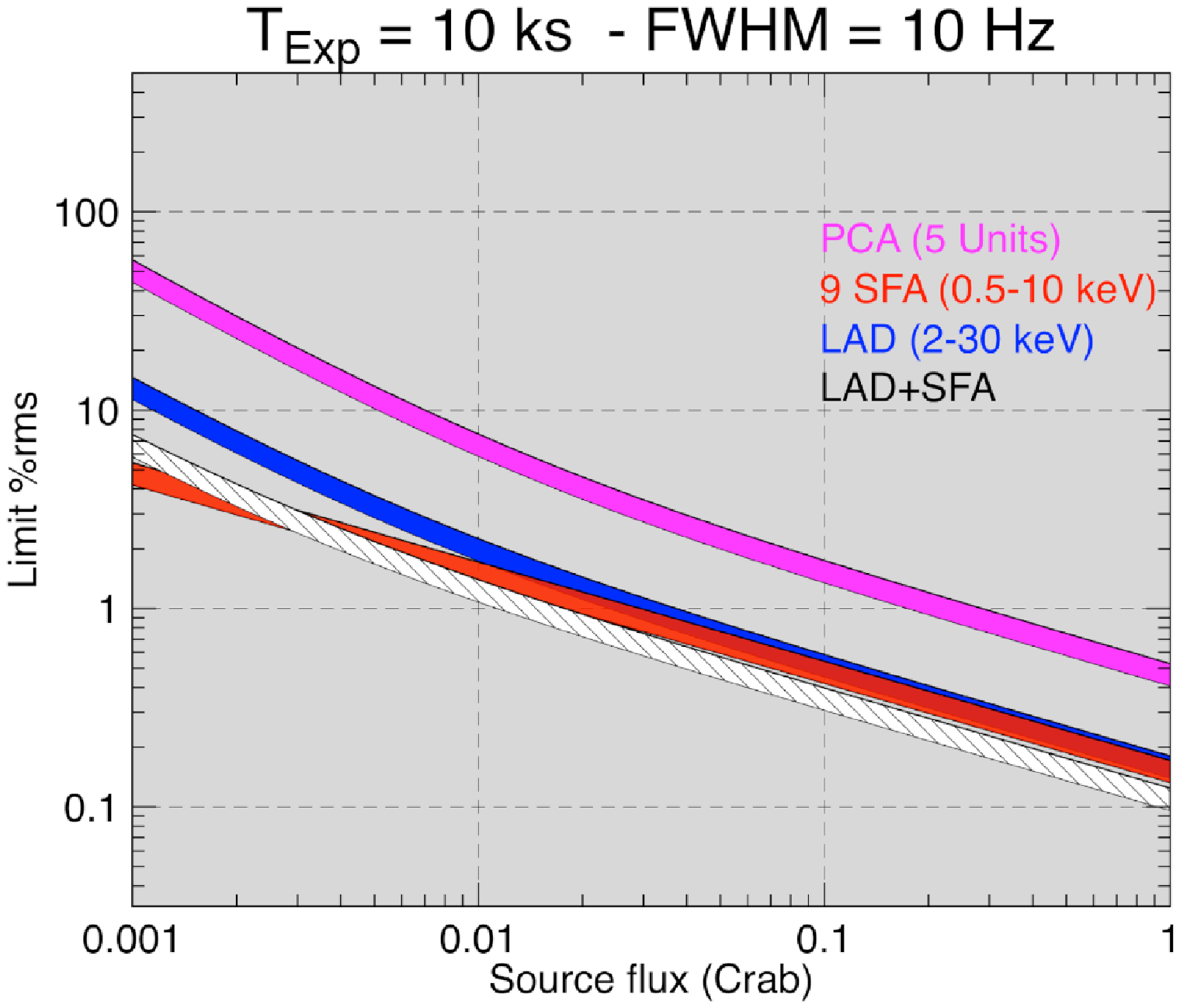}
\includegraphics[width=0.35\textwidth,angle=-90]{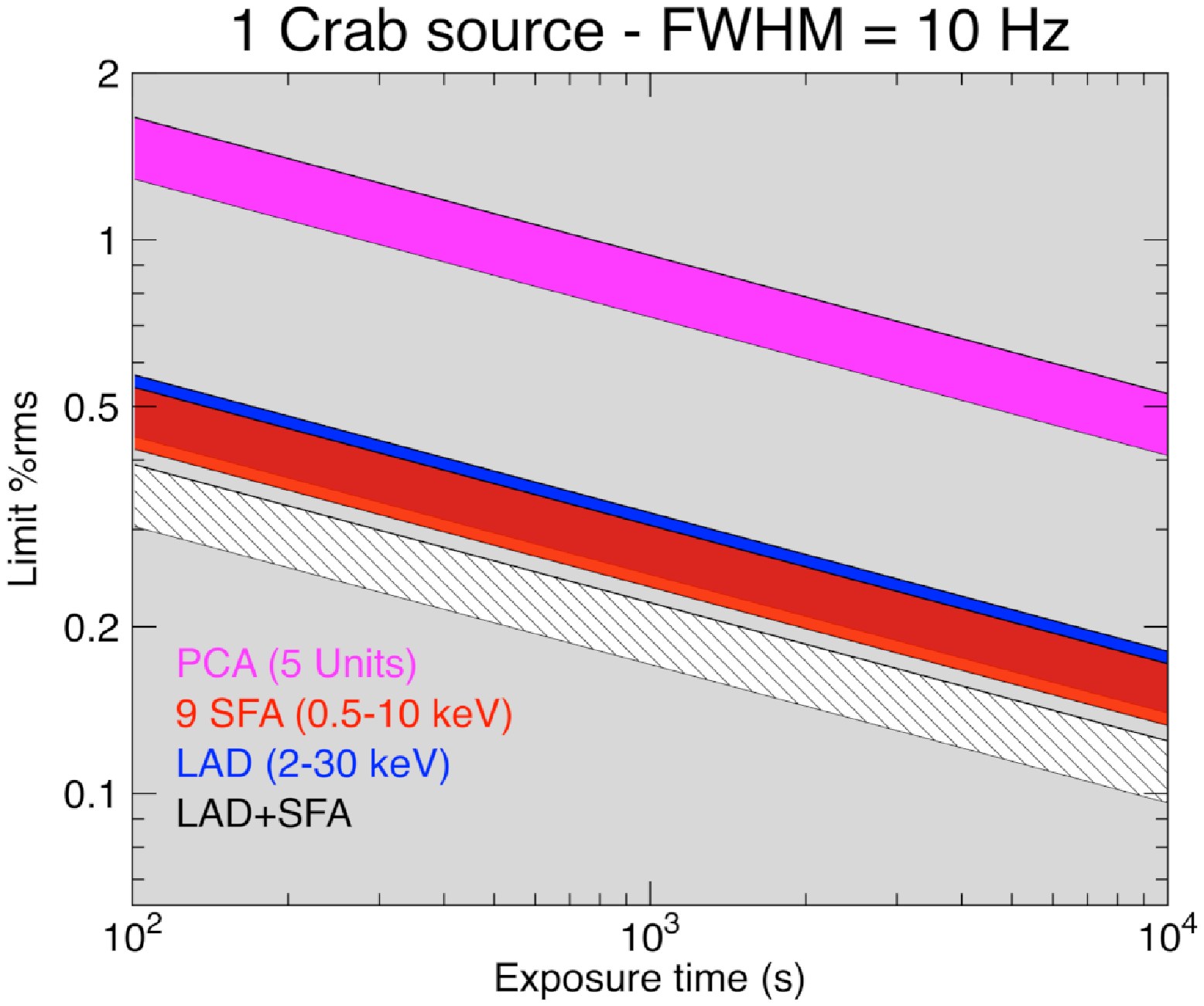}
\caption{\textit{Left.} Simulated {\it eXTP} sensitivity (for detection against the Poisson noise level) in fractional rms, for a QPO of arbitrary frequency with FWHM=10 Hz, as a function of flux for 10~ks exposure.  \textit{Right.} Simulated {\it eXTP} sensitivity in fractional rms, for a QPO of arbitrary frequency with FWHM=10 Hz, as a function of exposure for a source with flux equal to 1 Crab. In both panels each stripe marks the 3 sigma and 5 sigma significance levels (lower and upper boundary, respectively). Different colors correspond to different instrument: {\it RXTE} PCA (with 5 proportional counter units operating) (magenta); SFA, 9 telescopes (red); LAD, 40 elements (blue), sum of SFA (0.5-10 keV) and LAD (0.5-30 keV) (white).}
\label{fig:mottaqposim}
 \end{figure*}
 
High frequency QPOs (HFQPOs) are one of the most important discoveries made by the {\it Rossi X-ray Timing Explorer} ({\it RXTE}) (see ref.~\cite{vanderKlis2006}). They are normally found at several hundreds of Hz in black hole XRBs and appear stable in frequency at values that, as predicted for relativistic frequencies, scale inversely with black-hole mass \cite{ref48}. The black hole HFQPOs are particularly intriguing, but weak and transient, so it has been impossible to determine if their frequencies are really fixed, or only appear to be so because we are only just detecting them when they are strongest. Their amplitude distribution is severely cut-off by current instrumental limitations \cite{ref49}, but this will be remedied by {\it eXTP}.

~\cref{fig:mottaqposim} shows the sensitivity (in fractional rms) of the {\it eXTP} instruments (LAD, SFA and the combination of the two) for detection against the Poisson noise level of a QPO of arbitrary frequency and with a FWHM of 10 Hz. The left panel shows the case of a fixed exposure time of 10~ks for a variable source flux (between 1~mCrab and 1~Crab, the flux range where most XRBs are found during their active phases). The right panel shows the case for a source at 1 Crab flux, for a variable exposure time (100~s to 10~ks).  {\it eXTP} will bring a significant improvement in HFQPO sensitivity compared to the {\it RXTE} Proportional Counter Array (PCA), the only instrument to detect HFQPOs to date.  This improvement will allow the detection in short intervals of transient HFQPO signals, enabling us to study the duty cycles of the HFQPO signal and its links to changes in inner accretion structure which will be measured simultaneously by the suite of other techniques (e.g. reflection spectroscopy, reverberation) described in this paper.

\begin{figure*}
\centering
\includegraphics[scale=0.7,angle=90]{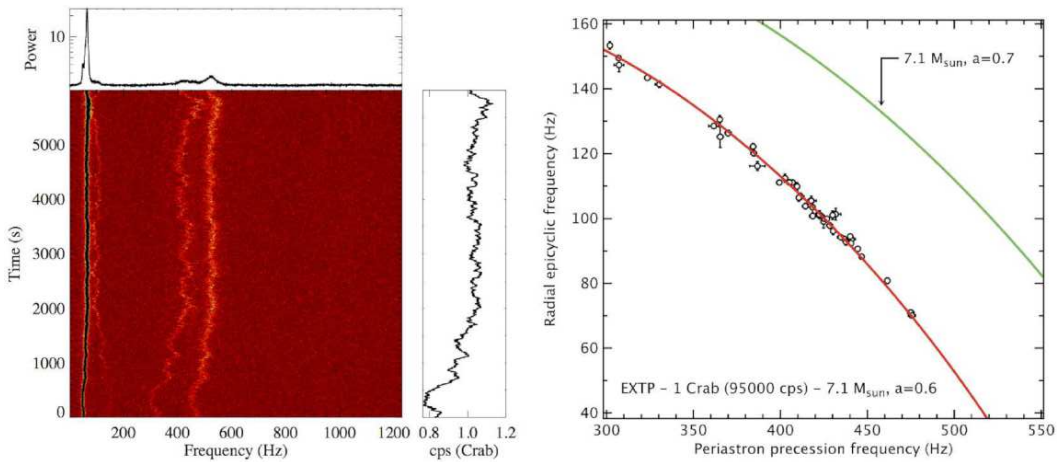}
\caption{\textit{Left.} Simulated dynamical power spectrum of HFQPO signals from a 1 Crab BH XRB observed with {\it eXTP} LAD for 4 ks.  The simulation assumes the RPM model for the QPO and includes (from low to high frequency) the contributions from Lense-Thirring precession, radial epicyclic motion, periastron precession and orbital motion.  The QPO frequencies drift as the disk inner radius changes (leading also to correlated flux variability).  \textit{Right.} Comparison of radial epicyclic and periastron precession frequencies measured using maximum likelihood fitting \cite{ref54} of power spectra from 100~s segments of the same simulation, including as a solid line the expected curve for the black hole mass (7.1~M$_{\odot}$) and spin ($a=0.6$) assumed in the simulation.  Measuring such a curve will be a powerful confirmation of the RPM model.  The curve for $a=0.7$ is also shown, demonstrating that {\it eXTP} can use the HFQPOs to place tight constraints on spin, independent of other methods or the assumption that the disk inner radius is fixed at the ISCO.}
\label{fig:barretdynpow}
\end{figure*}

This leap forward in HFQPO detection capability will enable a further leap in our understanding of the physical origin of the HFQPO signal, potentially allowing detection of weak signals associated with the epicyclic motion expected in strong-field gravity, along with accurate measurement of the key parameters of the Kerr metric: black hole mass and spin.  Given their rarity and the detection limits in {\it RXTE} data, it is likely that the relatively few HFQPOs observed to date are the `tip of the iceberg' of a population of weaker signals which may occur simultaneously at multiple frequencies, as predicted by the best current models for the QPOs as being linked to orbital motion in strong-field gravity.  
For example, besides the expected strong signal linked to the orbital motion and Lense-Thirring precession, which produce respectively the main high and low-frequency QPO peaks, the relativistic precession model (RPM, \cite{stellaetal1999}) explains additional weaker high-frequency signals as being due to GR effects, notably radial epicyclic motion (for which QPO candidates have not been detected to date) and associated periastron precession (which in the RPM can explain the lower-frequency signals in pairs of HFQPOs).
These signals have been poorly-studied to date, due to limitations in data-quality, but could be detected in as little as 100~s by {\it eXTP} and are expected to evolve significantly with changes in the inner accretion flow, e.g. linked to changing flux, in a way which can be detected within just a few ks using a {\it dynamical power spectrum} of the photon count rate (see ~\cref{fig:barretdynpow}, left panel).
In the Kerr metric, the radial epicyclic and periastron precession frequencies are uniquely related (via the orbital frequency) to the black hole mass, radius and spin, such that if the inner disk radius changes (as envisaged by the RPM model), the comparison of all three frequencies will provide a powerful test of the model.  If the RPM model is confirmed, measurements of multiple, changing frequencies for any pair of these high-frequency signals may be then be used as a powerful spin and mass estimator. For example, splitting the {\it eXTP} simulated RPM signal into 100~s segments to accurately measure both radial epicyclic and periastron precession QPO frequencies, we show the comparison in ~\cref{fig:barretdynpow} (right panel), where a strong distinction is seen between the `observed' $M_{\rm BH} = 7.1$~M$_{\odot}$, $a=0.6$ curve and a curve for the same mass but spin $a=0.7$.  Such measurements could also be made using the low-frequency QPO, which in the RPM model corresponds to nodal (Lense-Thirring) precession (see below).  However, since other, accretion-related effects may influence these lower-frequency signals \cite{ref52}, the HFQPO measurements should be seen as the strongest test.  It is important to note that a HFQPO measure of spin would be completely independent of the other spin-measurement techniques described in this paper, yielding a further powerful test on these measurements, should the RPM model turn out to be the correct model for HFQPOs.  If the HFQPOs are better explained by other models, different frequency behaviour will be highlighted in the data.  For example, the epicyclic resonance model \cite{ref50} predicts HFQPOs at a constant resonant frequency, regardless of changes in inner disk radius, with combinations of additional frequencies which yield their own checks on the spacetime metric.

Low frequency QPOs (LFQPOs, with frequencies below $\sim50$~Hz in black hole systems) have been known for many years and are divided into different types according to their phenomenological properties \cite{Wijnandsetal1999,remillardetal2002,ref51}.  The variations in the characteristic frequencies of the Type-C QPOs are associated with the hard states and are likely related to variations of the inner disk truncation radius (see ~\cite{ref52,ref53}), offering the possibility to directly track the changes in the geometry of the accretion flow. On the other hand, the abrupt appearance/disappearance of the so-called type-B QPO might be the X-ray signature of the occurrence of fast relativistic ejections along the jet \cite{ref55,ref56}.  

Several models have been proposed for these QPOs in terms of strong-field orbital and epicyclic motion in the disk flow, most notably the `Lense-Thirring' precession \cite{ref62} of the inner hot, geometrically thick accretion flow caused by GR frame-dragging of orbiting plasma due to the misalignment of the black hole spin and the angular momentum of the accreting material in the disk.  Interpretations along these lines have received support from large-scale MHD simulations \cite{ref63,ref57,Liskaetal2017}.  Precession models have also gained significant observational support in recent years from the observed dependence of key type-C QPO properties (rms amplitude and energy-dependent phase lag) on the inclination of the X-ray binary system orbit, with higher-inclination systems showing systematically higher rms amplitudes and soft lags as opposed to the low rms amplitudes and hard lags seen in lower-inclination systems \cite{ref56,ref58}.  Such behaviour is easy to explain when these QPO characteristics are linked to a changing emission geometry as seen by the observer (as expected from precession of the inner flow), as opposed to an intrinsic variation in accretion rate.

LFQPOs, thanks to their intrinsically high amplitudes, are normally detected with high significance even by instruments with limited sensitivity.  With {\it eXTP} observations, LFQPOs will become a much more effective tool for the study of strong-field gravity.  Firstly, the greatly increased sensitivity to high-frequency signals will enable the LFQPOs to be detected simultaneously with higher-frequency signals. These new measurements will allow dynamical frequencies and the slower, possibly precession frequencies to be combined, to give strong constraints on black hole mass and spin \cite{Mottaetal2014a,Mottaetal2014b}.  Secondly, the high count rates obtained by {\it eXTP} will routinely allow the detailed study of the QPO waveform and/or the dependence on the QPO phase of the spectral shape of XRBs (tomography). These techniques will enable a new form of mapping of the inner accretion flow and the changing geometry associated with the QPO, which will be discussed in Sect.~\ref{ssect:QPOtomo} and \ref{ssect:QPOpol}.

\section{Mapping the inner regions}
\label{sect:inner_region}

The methods described in the previous Section make use of either `time-averaged' (spectral fitting and polarimetric modelling) or `energy-averaged' (power-spectral) methods to obtain some key diagnostics of matter behaviour in strong field gravity.  Such methods will always suffer to some extent from systematics made by the implicit time or spectral averaging that is done on what are complex, variable, multi-time-scale and multi-component processes, leading to potential model degeneracies.  However, the large collecting area, good spectral capability and high time-resolution and count rate capability of {\it eXTP} allows a new approach, currently applied in a few cases and still being developed, of combining spectral, timing and (in the future) polarimetric information.  These spectral-timing(-polarimetry) methods enable us to disentangle different spectral and polarization components according to their variability and causal relationships, either in phase (for quasi-periodic variations) or time-delay with respect to one another (for aperiodic broadband noise variability).  Thereby we can significantly reduce the \textit{systematic} error inherent from the degeneracy involved with time- or energy-averaged modelling.
In the following, we report on detailed simulations based on spectral-timing and spectral-timing-polarimetric techniques, applied to \textit{eXTP} data from both XRBs and AGN. 
These studies are necessarily preliminary, as the techniques and models to describe what is a new kind of data are still under development.  But it is already clear that these methods combined with {\it eXTP} data will provide an innovative and powerful way to map the accretion flows and emitting regions in the strong field gravity regime of accreting black holes.

\subsection{Accretion flow tomography with LFQPOs}
\label{ssect:QPOtomo}
\begin{figure*}[t!]
\centering
\includegraphics[scale=0.6]{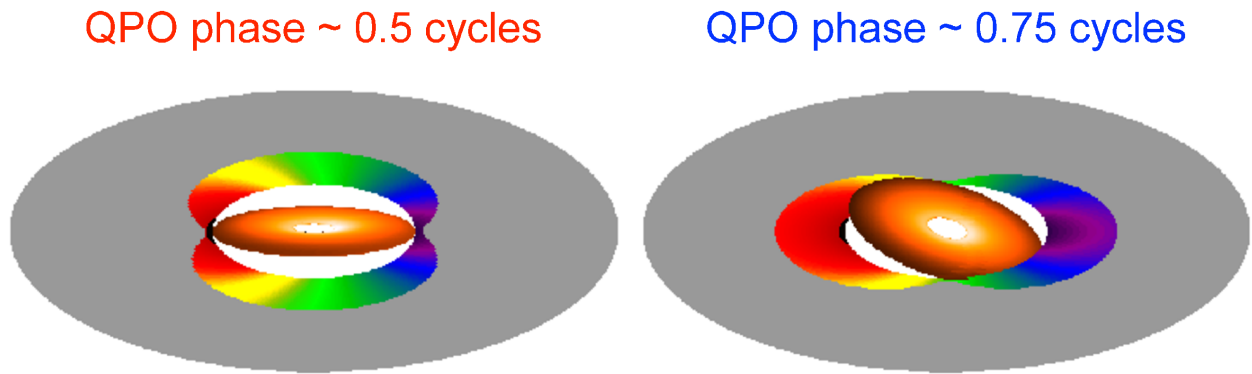}
\caption{Schematic geometry for a precessing inner hot flow at two phases of the resulting QPO signal, showing the different azimuthal regions on the disk (coloured regions on grey) that are predominantly illuminated by the top and bottom sides of the flow.  In this example we assume that the hot flow is seen more edge-on at a phase of 0.5 cycles, and that orbital motion of the disk is clockwise as seen on the page.  The colour scheme indicates whether disk emission is redshifted or blueshifted due to relativistic effects. The relativistic distortion of the disk reflection spectrum should change systematically with QPO phase.}
\label{fig:qpogeom}
\end{figure*}

The high throughput and spectroscopic and polarimetric capabilities of {\it eXTP} make it uniquely suited to probing the physical origin of LFQPOs and using the QPO signal itself as a probe of the inner flow structure and the effects of strong-field gravity.  For example, in the case of the strongly-favoured precession models for the QPO, as the inner hot flow precesses it should illuminate different azimuths of the accretion disk, leading to a variation in the appearance of relativistic reflection signatures (\cref{fig:qpogeom}).  The rapid orbital motion of the disk material means that an inclined observer will see a blue-shifted line when the approaching material is illuminated and a red-shifted line when the receding material is illuminated \cite{IngramDone2012FeK}.  Thus, phase-resolved spectroscopy of a QPO from a precessing hot inner flow in an XRB enables {\it tomography} of the disk emission.  The predicted quasi-periodic modulation of the iron line centroid energy was recently detected ($3.7 \sigma$ significance) for the first time using {\it XMM-Newton} and {\it NuSTAR} data from the black hole XRB H 1743-322 \cite{ref61}, and tomographic mapping was subsequently carried out \cite{Ingrametal2017}. This provided an excellent proof of principle demonstration of the technique, but required a very long exposure of a comparatively dim source so as to avoid problems of photon pile-up in the {\it XMM-Newton} detectors.

\begin{figure*}[t!]
\centering
\includegraphics[scale=0.8]{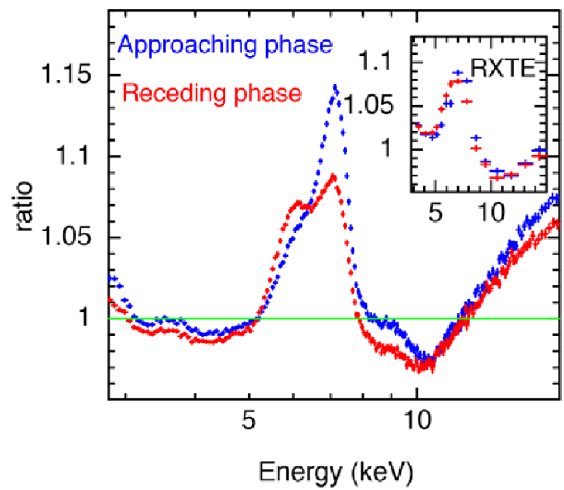}
\caption{Simulation of a $50$ ks {\it eXTP} observation of GRS 1915+105,
  split up into 20 QPO phase bins, each of $2.5$ ks integration time. 
  We assume a disk truncation radius of $30~r_g$, a disk inclination angle of 
  $70^\circ$ and an empirical, phase-dependent illumination pattern on 
  the disk to match that obtained for H~1743-322 by \cite{Ingrametal2017}. 
  The synthetic data are plotted as a ratio to
an absorbed power-law with photon index $\Gamma=1.8$.The main
panel shows the synthetic LAD data for the QPO phase bins
corresponding to maximum illumination of the approaching disk material
(blue) and maximum illumination of the receding material (red).  
The change in line profile between QPO phases is
clearly resolved by the LAD. The inset shows the same simulation for
the case of {\it RXTE}, which fails to resolve the change in line profile, due to photon
counting noise and the modest spectral resolution offered by the {\it RXTE} PCA instrument.}
\label{fig:fedance}
\end{figure*}

The large-area detectors of {\it eXTP} enable phase-resolved spectroscopy of any of the LFQPOs observed to date in black hole XRBs, by applying state-of-the-art techniques to recover the QPO phase from the Fourier information on short time-scales and bin the spectral data accordingly \cite{ref61,ref64}. ~\cref{fig:fedance} shows LAD data from a simulated {\it eXTP} observation of GRS~1915+105 in an intermediate state. The assumed spectrum is modelled to agree with an \textit{RXTE} observation of the source from 6$^{\rm  th}$ March 2002, when it exhibited a strong QPO at $0.46$ Hz and prominent iron line.  The very high
count rate detected by the LAD means that the change in shape of the iron line as a function of QPO phase resulting from Lense-Thirring precession of the inner flow can be clearly detected. It is not only possible to detect that the iron line has a higher centroid energy during the approaching phase, but it is also clear that the shape of the red wing ($\sim 6$ keV) and the smeared iron K edge ($\sim 9$~keV) changes significantly between QPO phases. The inset shows the same simulation for the PCA instrument on board \textit{RXTE}.  For the real \textit{RXTE} observation that this simulation is based on, Ingram \& van der Klis \cite{ref64} did find tentative evidence of a line centroid energy modulation, but only with a significance of $1.9\sigma$.

Other current observatories can potentially improve upon the current state-of-the-art, but without the large step-change in performance expected for {\it eXTP}. For example, the \textit{XMM-Newton} EPIC-pn instrument and \textit{NuSTAR} mission all provide the required spectral resolution, but have lower effective area at the iron line than \textit{RXTE} (while \textit{XMM-Newton} cannot observe such bright sources due to pile-up and telemetry issues and \textit{NuSTAR} has its count rate limited by instrumental deadtime). \textit{ASTROSAT}'s LAXPC instrument \cite{Antiaetal2017} has a comparable area and energy resolution at Fe~K energies to \textit{RXTE}.  The recently launched \textit{Neutron Star Interior Composition Explorer} (\textit{NICER}, \cite{Gendreauetal2016}) has CCD spectral resolution and high count-rate capability, but also has a soft spectral response, covering 0.2-10 keV and peaking at 1.5~keV, which is not so well-matched to the hard shape of the QPO spectrum, so is not optimal for studies of reflection modulation.

The LAD allows a vast improvement on all of these alternatives, with the required 
spectral resolution and far larger effective area at Fe K energies ($\sim6$ 
times greater than \textit{ASTROSAT} LAXPC and $>40$ times greater than instruments 
such as {\it XMM-Newton} EPIC-pn and {\it NICER} which have comparable CCD-like 
spectral resolution to the LAD), as demonstrated by ~\cref{fig:fedance}. 
How the iron line changes shape with QPO phase depends on how the disk illumination profile evolves with precession phase, which ultimately depends on the geometry and emission mechanism of the inner flow. The phase-resolved iron line profile also depends strongly on the disk inner radius, which sets the rotational velocity of the reflecting material, and the inclination, which sets the line-of-sight velocities. These parameters can be measured accurately with high quality data from the LAD. Fitting a precession model \cite{Ingrametal2017} to the synthetic LAD data shown in ~\cref{fig:fedance} gives an inclination angle of $i=66.7^{+2.3}_{-2}$ degrees and a disk truncation radius of $27.2^{+1.5}_{-1.5}~r_g$. We can achieve even greater precision by considering all 20 QPO phases rather than just two, and also including the data from the SFA.  Observing how the spectrum evolves with QPO phase provides many more independent data points than the time-averaged spectrum, with each phase preferentially sampling the reflection spectrum from different azimuths on the disk, which (as seen by the observer) experience different relativistic effects.  E.g. emission from the sides moving perpendicular to our line of sight is mostly affected by gravitational and special-relativistic time-dilation, while emission from the approaching side also undergoes substantial Doppler boosting.  This differential sampling of relativistic effects allows model parameter degeneracies to be more easily broken.

\subsection{Accretion flow geometry with QPO timing-polarimetry}
\label{ssect:QPOpol}
\begin{figure*}[ht!]
\centering
\includegraphics[width=0.8\textwidth,angle=0]{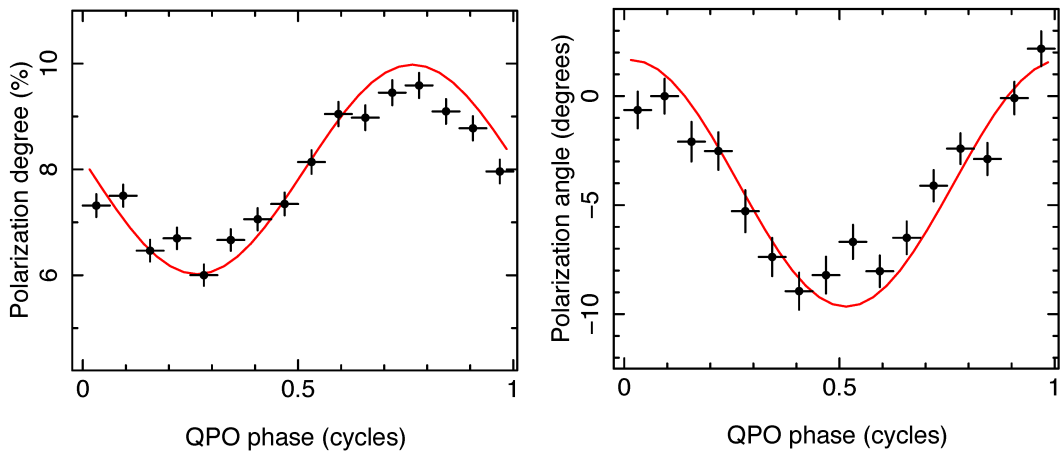}
\caption{Variation of observed coronal polarization due to quasi-periodic precession of the corona: measured (black points) and input (red lines) polarization degree (left) and angle (right) as a function of QPO phase from our simulation of a 50~ks exposure of GRS~1915+105 in a state showing a 1 Hz QPO. We see that the variations in polarization properties are recovered well by {\it eXTP}.}
\label{fig:qporesults}
\end{figure*}

In addition to measuring changes in the iron line profile with QPO phase, {\it eXTP} will be able to measure
changes in polarization degree and angle with QPO phase, giving an additional powerful and independent probe
of the variations in coronal geometry associated with the QPO modulation (see also discussion in Sect. \ref{ssect:xrb coronae}). We simulate this variation in polarization properties for the same synthetic GRS 1915+105 observation described above. According to the calculations of \cite{ref65}, Lense-Thirring precession of a flattened hot inner flow will also produce a QPO in the polarization degree with an absolute rms amplitude (at the fundamental frequency) of $\sim 1.4\%$ and a QPO in the polarization angle with absolute rms amplitude (again, at the fundamental frequency) of $\sim 4^\circ$. 

We simulate a 1 Hz QPO in the flux and polarization properties, taking into account the quasi-periodicity of the oscillation and also the coincident broadband noise observed in the flux, which is intrinsic to the source rather than instrumental. The QPO phase drifts on a random walk away from that of a strictly periodic sine wave, as is observed for QPOs in GRS 1915+105 \cite{ref66}. The flux, polarization degree and polarization angle vary sinusoidally with this varying QPO phase, following the results of \cite{ref65}.   Data corresponding to these variations are then simulated accounting for the detector characteristics of the SFA and LAD.  Here we have used the high count rate ($\sim 70000$~count~s$^{-1}$) LAD synthetic light curve to assign instantaneous QPO phase values through a filtering method \cite{vandenEijnden2016}. Using the phase values assigned from the synthetic LAD light curve, we bin the Stokes parameters measured by the SFA into 16 phase bins and used them to calculate the polarization degree and angle for each bin. \cref{fig:qporesults} (black points) shows the simulated phase-folded polarization degree (left) and angle (right) plotted as a function of QPO phase. The points are clearly not consistent with constant polarization properties. The red line on each plot shows the input variation, which the simulation recovers well. 

{\it eXTP}'s combination of high throughput and polarimetric capabilities will  enable the search for HFQPOs in the polarization properties of black hole and neutron star XRBs. High frequency variations in the polarization degree and angle are expected for most HFQPO models, for example the orbiting hot spot model \cite{Schnittman2005,Beheshtipouretal2016}. Detection of such variations will provide a completely new way to diagnose the true HFQPO mechanism, and will also enable the HFQPO polarization signature to be used as a tool to map the inner accretion flow. 

We simulate a 50~ks observation of the upper HFQPO in the black hole XRB GRO J1655$-$40 studied by \cite{Mottaetal2014a} (i.e. a frequency of $441$ Hz, an rms of $4.5\%$ and a FWHM of $30$ Hz). We input sinusoidal variations of the polarization degree and angle as a function of HFQPO phase. For the polarization degree, we input a mean and standard deviation of $4\%$ and $1\%$ respectively and for the polarization angle, we assume a mean of $0^\circ$ and a variability amplitude of $10^\circ$, as expected from orbiting hotspot models (see Figure 15 of \cite{Beheshtipouretal2016} for comparison). Since detection of these high frequency features is more challenging, we use the Fourier method of \cite{IngramMaccarone2017}, which is more sensitive than the phase-folding method, making use of the cross-spectral combination of the polarisation signal from the SFA with the high-count rate light curve from the LAD, which is used as a reference signal to isolate the HFQPO polarisation signal. 

\begin{figure}[H]
\includegraphics[scale=0.35]{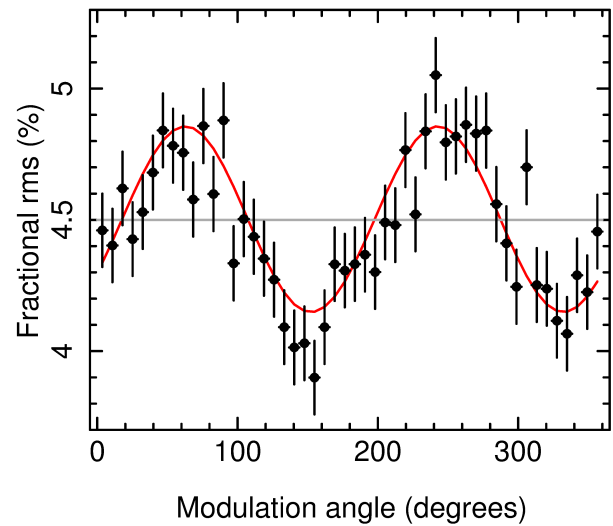}
\caption{Fractional rms of a simulated HFQPO as a function of modulation angle (which defines the direction of the electron track caused by a photon hitting the detector). The red line shows the sinusoidal pattern caused by quasi-periodic variability of the polarisation degree and angle at the frequency of the HFQPO (441 Hz in this case), and the grey line represents the null hypothesis in which only the flux varies with QPO phase, whereas the polarisation properties remain constant. The data points show the results from a simulated $50$ ks observation with {\it eXTP}, showing that the expected modulation is easily detected using the SFA/LAD combination.}
\label{fig:hfqpopol}
\end{figure}

Figure \ref{fig:hfqpopol} shows the resulting sinusoidal modulation of fractional rms at the HFQPO frequency as a function of detector modulation angle (red line), and the null-hypothesis of constant polarization properties (grey line). The data points show the recovered modulation from the simulated {\it eXTP} data.  Such a detection of HFQPO polarisation would not be possible with currently planned single-instrument X-ray polarimeters such as on board the {\it Imaging X-ray Polarimetry Explorer} ({\it IXPE}). The larger area of the {\it eXTP} PFA is an advantage here, but the most pivotal factor is the photon collecting capability of the LAD. The high count rates collected by the LAD  enable polarimetric-timing to be conducted for high frequencies ($>100$ Hz), commensurate with the azimuthal epicyclic frequencies of GR in the vicinity of the compact object and allowing the strong light-bending effects on the polarized signal in these regions to be studied.

Individually, the QPO phase dependence of both the iron line profile and the polarization properties, provide powerful diagnostics of the accretion flow geometry. When analysed jointly, as will be uniquely possible with {\it eXTP}, it will be possible to piece together the detailed geometry of the emitting region closest to the black hole.


\subsection{AGN Doppler tomography of orbiting hot spot patterns}
\label{ssect:AGNtomo}

\begin{figure*}[t]
\centering
\includegraphics[width=0.37\textwidth,angle=-90]{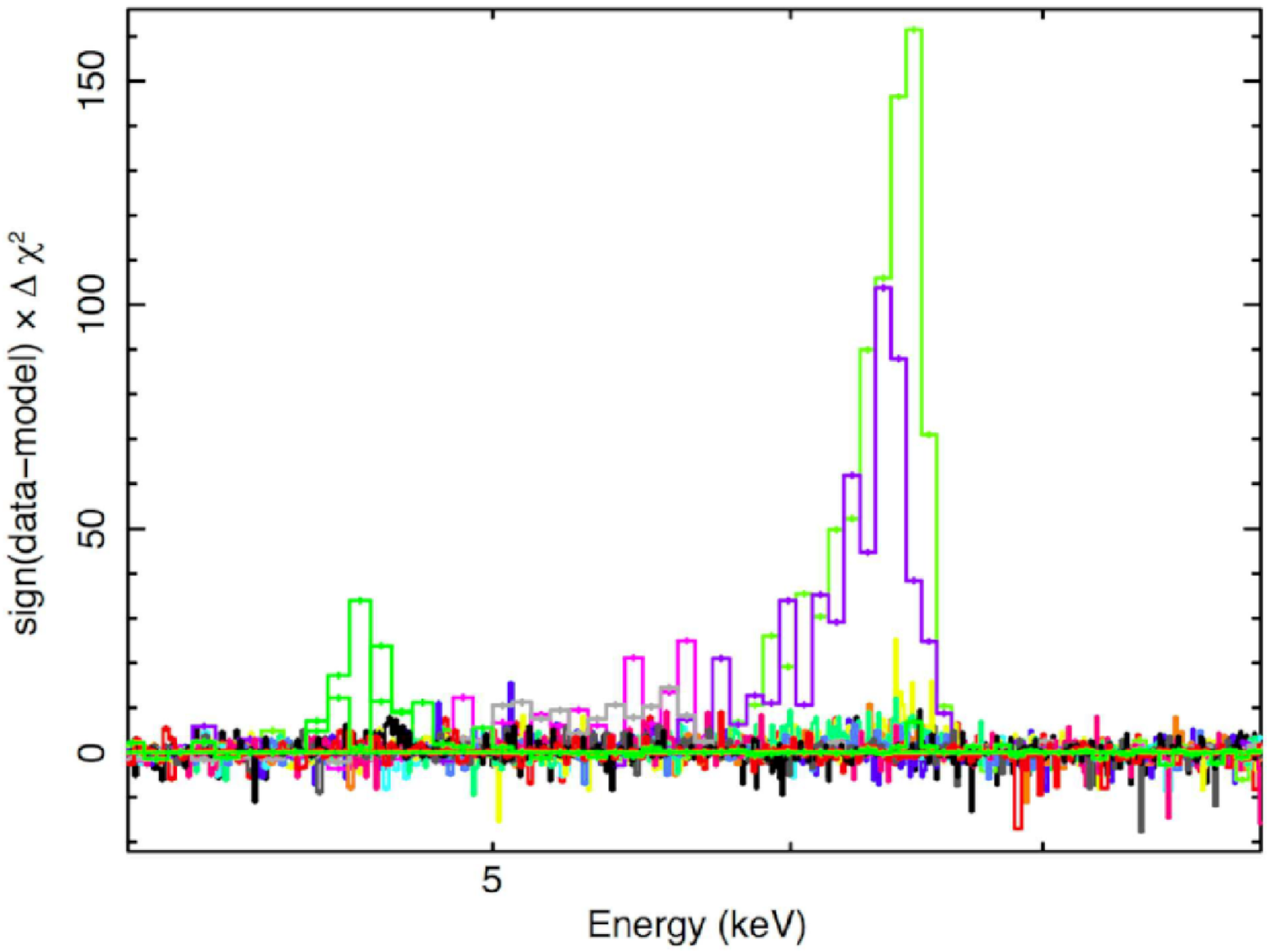}
\includegraphics[width=0.37\textwidth,angle=-90]{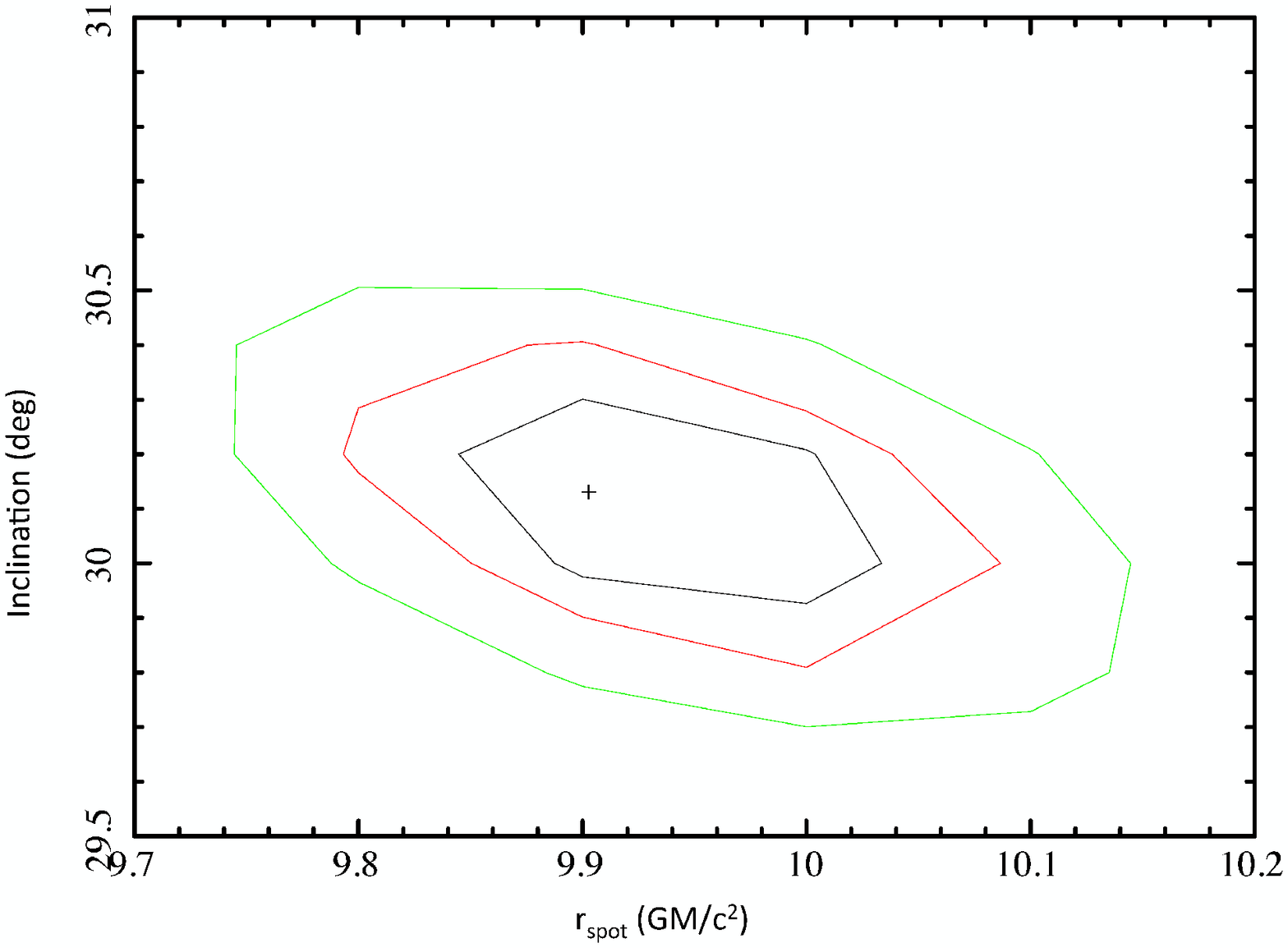}
\caption{AGN Doppler tomography. \textit{Left:} {\it eXTP} (both LAD and SFA) Fe line profile residuals (in terms of contributions to chi-squared) to line average resulting from two 10-ks orbits of a coronal hot spot around a 10$^{7}$ M$_{\odot}$ spin $a$=0.5 black hole at 10 $r_{g}$, for a total of 20 ks (~3000 s per profile plotted, each color represents a different phase on the accretion disk). The assumed disk inclination is 30$^{\circ}$. \textit{Right:} Error contours (1, 2  and 3 sigma in black, red, and green respectively) of disk inclination vs. hotspot orbital radius resulting from a fit to the line profiles at three different phases.}\label{fig:AGN_tomo}
\end{figure*}

In AGN the characteristic time scales of variability are much longer with respect to BH XRBs (see Sect. \ref{sect:section1}) allowing us to collect a larger number of photons per cycle. Despite the fact that this characteristic represents a great opportunity to investigate the physical origin of QPOs, we have only one convincing detection of AGN QPO in the Narrow Line Seyfert (NLSy1) RE~J1034+396 (\cite{ref9}). 

As the inhomogeneities associated with QPOs inevitably lead to variations in patterns of emissivity and illumination of the disk (see Sect.~\ref{ssect:QPOtomo}), the variations in emissivity patterns can also occur due to orbiting inhomogeneities in the disk due to `hot spots' \cite{dovciaketal04}.
These inhomogeneities can be generated if  the  disk  illumination  is provided by a localized flare just above the disk, rather than a central illuminator or an extended corona. 
Orbiting hot spot patterns in the accretion disk should undergo alternating Doppler redshifts and blueshifts, which lead to quasi-periodic distortions of the spectrum, including the broad Fe line profile.
The hotspot orbiting luminous blob in the accretion disk will cause a feature to move back and forth through the line profile as the blob transits the red- and blue-shifted regions. This signal can be used to reconstruct the geometry with the technique of Doppler tomography, as well as probing the orbital dynamics of the flow. 

Narrow and (apparently) transient features in the $\sim$ 4--6 keV energy range were observed in the  AGN NGC~3516 \cite{ref67,ref68} and in a larger XMM sample \cite{ref69}; they were characterized by relatively short-time-scale variability, of the order of tens of ks, but were not confirmed in subsequent observations, being at the limits of current X-ray telescope sensitivity.  More recently the double-peak structure in X-ray flares  observed in Sgr A$^\star$ has been reproduced with a simple orbiting hotspot model \cite{karssenetal17}. It has been argued that a hotspot can be stable long enough to create a flare. {\it eXTP} has the capability to confirm whether these intriguing features are real and if so, to use them as a powerful probe of the dynamics of matter in strong field gravity.

In AGN, where the orbital time-scale around the ISCO is of the order of a few ks (for 10$^7$M$_{\odot}$), {\it eXTP} will be able to follow the distortion of the Fe line profile due to orbiting hotspots in the time domain. As an example of the power of {\it eXTP} in applying this technique, we report in \cref{fig:AGN_tomo} (left panel) the  Fe line profile ratios to line average resulting from two 10-ks orbits of a coronal hot spot around a 10$^{7}$  M$_{\odot}$, spin $a$=0.5 black hole at 10 $r_g$ and contributing 10 per cent of the Fe line flux in a bright 2.5 mCrab AGN observed for a total of 20 ks (about 3000~s per profile plotted). Only the fluorescent spectral component is plotted \cite{ref21}. As these measurements rely on variability in the residuals, any narrow lines in the profile (arising by necessity at larger radii and hence varying much slower than the 10 ks inner disk orbital period, see Sect. \ref{ssect:astro_agn}) drop out automatically.

In the right panel we show the error contours (1, 2, and 3~$\sigma$ in black, red, and green, respectively) of disk inclination \textit{versus} hotspot orbital radius resulting from a fit to the line profiles in the three different phases. 
The hot spot orbital radius in $r_{g}$ can be measured to a precision of 1-2 per cent and combined with the spin measurement from the average line profile (see Sect. \ref{ssect:reflection}) allows measurement of the black hole mass to $<$30 per cent.  Conversely, if the black hole mass is independently measured (e.g. via optical reverberation mapping \cite{ref70}), spin can be determined to a precision comparable to that of the mass determination.  Such a constraint would apply to any hotspot radius and hence is independent of whether or not the disk is assumed to extend to the ISCO, as is the case for spin constraints from spectral fitting.  Thanks to its unprecedented throughput, {\it eXTP} will be able to perform such measurements for any AGN with flux above 1 mCrab.

\subsection{Measuring light-travel times to the inner disk} 
\label{ssect:reverberation}
\begin{figure}[H]
\includegraphics[scale=0.25]{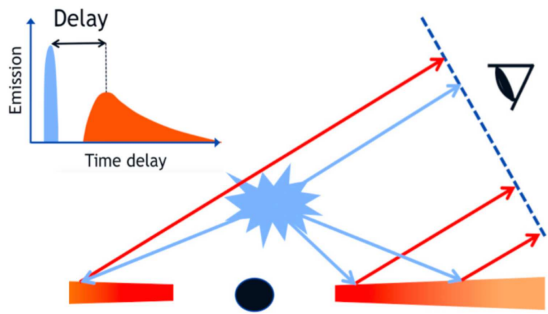}
\caption{Conceptual schematic of X-ray reverberation showing how the coronal emission is reflected by the disk resulting in a light-travel time delay of the reflected emission which constrains the light-travel time.  Note that in strong gravitational fields, strongly curved light-travel paths are expected (i.e. light-bending leading to the so-called `Shapiro delay') the effects of which can be constrained with accurate reverberation measurements.}
\label{fig:revcartoon}
\end{figure}

The X-ray power-law emission shows rapid aperiodic variability, on time-scales down to milliseconds for stellar mass black holes, and minutes for the supermassive black holes in AGN.  The fastest time-scales of the variability show that the bulk of the power-law emission originates in a central compact corona, which may be less than 10~$r_g$ in radius.  The corona illuminates the accretion disk and is reprocessed to produce the observed relativistically broadened reflection signatures (see Sect. \ref{ssect:reflection}) and (due to heating by the absorbed part of the incident flux) extra blackbody emission, which is hot enough to be emitted in the X-ray band in stellar mass black holes and also in high accretion-rate, low black hole mass AGN.  However, the reprocessed emission does not respond instantly to coronal variations: it is delayed with respect to the observed power-law variations by the extra light-travel time from the corona to the disk and then to the observer (see Figure~\ref{fig:revcartoon}).  This effect is known as {\it reverberation} and, due to the compactness of the emitting region and the fact that the delay is simply related to a light-travel time, it can be used to {\it reverberation map} the innermost regions of accreting black holes on scales down to the event horizon \cite{ref71}.    

Unlike conventional spectral-fitting, which reveals only velocity shifts and strong-field gravity effects (i.e. Fe line redshifts), from which radii are inferred in relative units of the gravitational radius (which assumes relativistic motions in a known metric), reverberation mapping also reveals distances of emission in absolute units (i.e. km) given by the light travel time.  Thus it provides an {\it independent and complementary check} on the emission geometry inferred from spectral-fitting techniques. In combination with the dynamics obtained by spectral-fitting measurements, reverberation can determine the mass of the central object \cite{ref20}.  In other words, the optical reverberation methods currently routinely used to measure AGN black hole masses using optical line emission from thousands of $r_g$ \cite{ref70} can be applied on much smaller size-scales of a few $r_g$ (and to stellar mass black holes).  In cases where a black hole mass is already known (e.g. from optical reverberation in AGN, or binary orbital dynamics in X-ray binary systems), it can be compared with the estimate based on X-ray reverberation of the innermost regions, to provide a powerful consistency test of black hole mass estimated from dynamics in the weaker field at thousands of $r_g$, to the strong-field at a few $r_g$.

To date, reflection reverberation signatures have been discovered in the short-time-scale variability of a number of AGN (e.g. see \cite{Fabianetal2009,Zoghbietal2012,Karaetal2016}), however the plots of lag vs. energy (`lag-energy spectra') used to detect reverberation via, e.g. local peaks in the lag at Fe~K energies, must be heavily binned in energy and hence have poor spectral resolution  (see ref.~\cite{ref71} for a detailed review).  Furthermore, the limits on making spectral-timing measurements of short-term variations in stellar mass systems mean that disk thermal reverberation has only been seen in a handful of X-ray binary systems \cite{Uttleyetal2011,DeMarcoetal2015} and Fe K reverberation is difficult to observe without very high count rates.  Thus we know that the X-ray reverberation phenomenon exists and that the observed lags are consistent with our basic physical picture, but we cannot yet confront them with detailed models for the innermost emitting regions, which can predict the energy-dependent time-delays and thus map those regions.  To make the step-change from phenomenology to mapping for both AGN and XRBs, we need much better signal-to-noise measurements at good (CCD-quality) spectral-resolution and across a broad energy range, so that we can study all the reverberation components simultaneously.  This is what {\it eXTP} will provide.
\begin{figure*}
\centering
\includegraphics[scale=0.8]{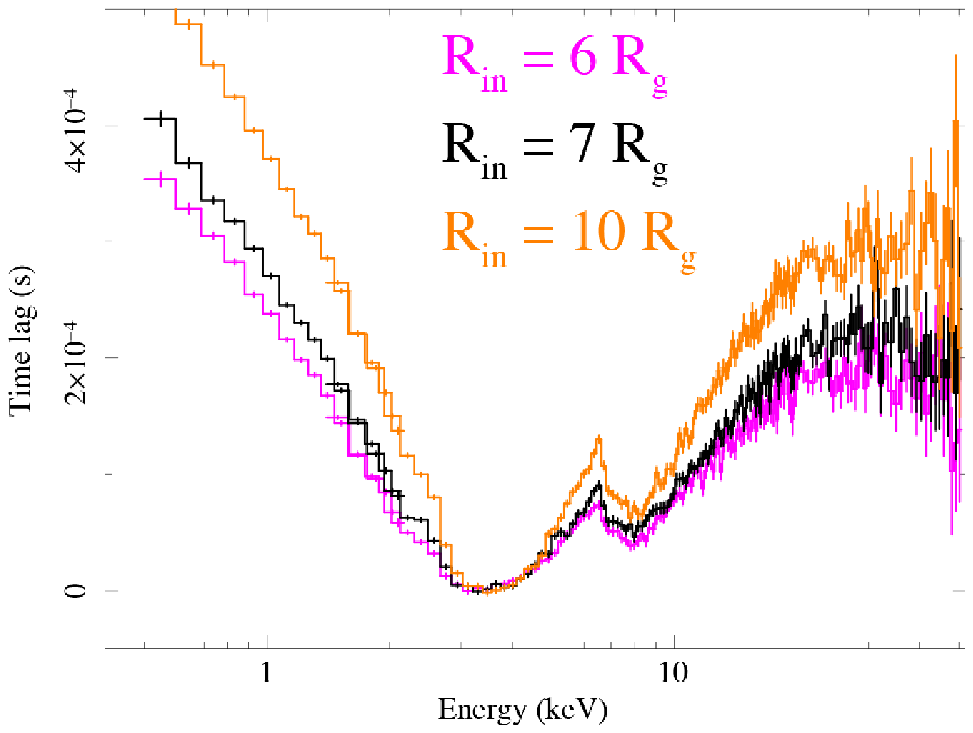}
\includegraphics[scale=1.05]{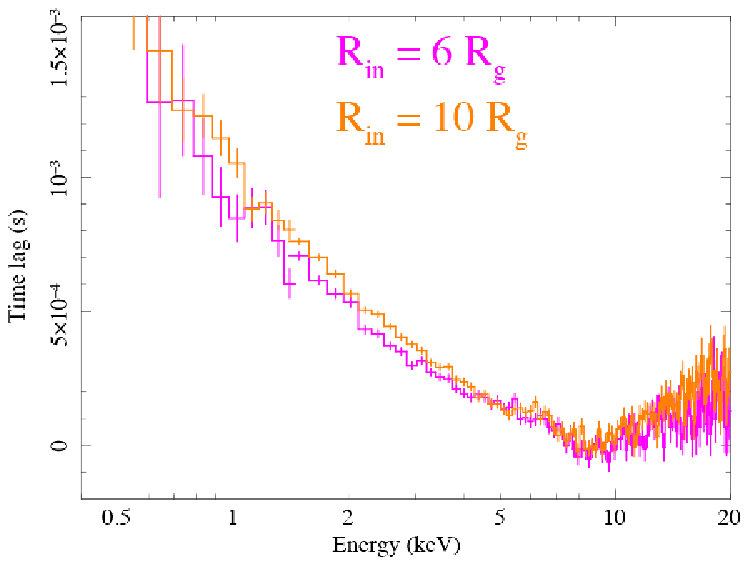}
\caption{Simulated {\it eXTP} (100~ks exposure) lag vs. energy spectrum of a 10 M$_\odot$ black hole XRB with 30~degrees disk inclination, for a bright (1 Crab) hard state ({\it left panel}), and a bright (3 Crab) soft-intermediate state ({\it right panel}).  The lags are obtained by integrating over the 50-150~Hz range the Fourier cross-spectrum obtained for each narrow energy bin (with respect to a broad 2--20~keV reference band, following the approach described in ref.~\cite{ref71}).  Different inner disk radii can be easily distinguished.  Thanks to the broad energy coverage of the SFA/LAD combination, the hard state lags can be accurately determined for the broad Fe K emission and reflection hump as well as the disk blackbody emission, allowing multiple independent measures of the geometry and relativistic effects on the innermost part of the accretion flow. In the disk-dominated soft-intermediate state, the lag at each energy can be used to determine the disk emitting region size at that energy and hence provides a test of our understanding of the distribution and radial temperature profile of disk emission.}
\label{fig:hardbhlags}
\end{figure*}

The signal-to-noise for spectral-timing measurements scales with the square-root of count rate for AGN but linearly with count rate for XRBs (see details in \cite{ref71}), so {\it eXTP} will enable the biggest improvements in reverberation measurements on the stellar-mass systems.  An example plot of lag vs. energy from a 100~ks {\it eXTP} observation of a black hole X-ray binary in a bright (1~Crab) hard state is shown in ~\cref{fig:hardbhlags} (left panel).  In the absence of a detailed understanding of coronal geometry, we assume a central point-like corona such that lags only include (for the different parts of the disk) the combination of radial light-travel distance to the corona plus the disk-to-observer distance.  Thus the simulations are not intended to be accurate predictions but rather illustrative of the strength of the lag signals due to different components and the relative lags between different energies, as well as the precision of the lag measurements.  Some of the remarkable features are:
\begin{itemize}
\setlength{\itemindent}{2em}
\setlength{\listparindent}{4em} 
\item[-] The broad band-pass of the SFA/LAD combination from 0.5 to 50 keV enables the reverberation lags of the disk blackbody, Fe K and Compton reflection humps to be measured simultaneously, so that each provides a separate measure of the disk inner radius (but all can be modelled simultaneously, for much higher accuracy as well as providing a powerful consistency-check).
\item[-] The CCD-quality energy resolution allows differences in line-shape to be easily measured in the light-travel lags, so that line redshift {\it and} location can be simultaneously combined to test the effects of GR at small radii, constrain the accretion disk dynamics {\it and} measure black hole mass. 
\item[-] {\it {\it eXTP}} can measure lags in response to much shorter-time-scale variability than previously accessible (in our simulated XRB examples, 50--150~Hz or $\sim10$~ms).  This capability prevents contamination of the lags by non-light-travel-time effects, such as viscous propagation of accretion fluctuations through the disk and corona \cite{Uttleyetal2011}.
\item[-] For a given coronal geometry, differences in disk inner radius of less than 5~per~cent can be easily measured.  Since the coronal geometry can itself be modelled using the lags and many other diagnostics provided by {\it eXTP}, this will allow an accurate and {\it independent} determination of the black hole spin (in addition to that obtained from standard spectral fitting or QPO timing), if the disk extends to the ISCO, which is likely as the source approaches the soft state.  If the disk is truncated to larger radii, the measurements provide an accurate probe of the accretion geometry to understand how the accretion flow changes as the source evolves through different accretion states with different outflow type and power.
\item[-] In BH XRB soft states the power-law is weak and the variability amplitude is small, making reverberation measurements difficult even with {\it eXTP}.  However, in the soft-intermediate states just prior to the transition to the soft state, {\it eXTP} will be sensitive enough to measure thermal reverberation of the disk in response to variability of the stronger power-law emission (see ~\cref{fig:hardbhlags}, right panel).  The capability of {\it eXTP} to measure the 
the time-delays of photons of different energies (corresponding to emission regions of different temperatures) will enable us to not only measure the disk inner radius but also compare the radial temperature profile with our physical expectation for standard accretion disks.  Thus, reverberation mapping of this state, which is just `on the edge' of the soft state, will provide an independent check on the assumptions used to obtain disk inner radius (and hence spin) using the disk continuum fitting method.  Furthermore, reverberation mapping of the disk thermal emission will allow us to trace the variations in the disk structure from the hard states, where it is likely to be truncated, through to the soft state where the continuum fitting approach may be applied.
\end{itemize}
\begin{figure}[H]
\includegraphics[scale=0.8]{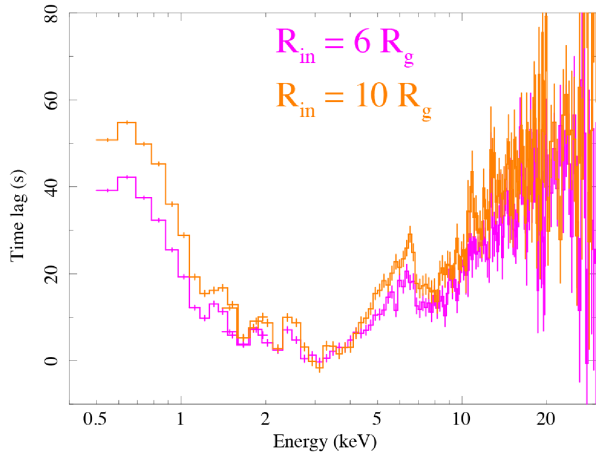}
\caption{Simulated lag-energy spectrum for a 100~ks observation of a 2~mCrab AGN with a $4\times 10^{6}$~M$_{\odot}$ SMBH, assuming a disk inclination of 30$^{\circ}$.  The lags are obtained by integrating the cross-spectrum of each energy (with respect to a broad 2--20~keV reference band) over the 0.3--3~mHz range.  Different radii can be easily distinguished and the broad SFA/LAD bandpass allows the lags of soft photoionized reflection to be accurately measured, as well as the broad iron line and Compton hump.}
\label{fig:agnlags}
\end{figure}
Although the improvements in AGN reverberation mapping with {\it eXTP} will not be as dramatic as for the XRBs, they will still be impressive compared to the current state-of-the-art, especially for bright AGN where the full {\it eXTP} bandpass (combining that SFA soft response with the LAD hard response) can be brought into use.  \cref{fig:agnlags} shows the results expected for a 100~ks {\it eXTP} observation of a 2 mCrab AGN containing a $4\times 10^{6}$~M$_{\odot}$ black hole, showing {\it eXTP}'s capability to simultaneously measure the lags associated with the soft ionised reflection, Fe line and hard reflection hump produced by reverberation close to the black hole.  Different inner radii can be easily distinguished, so that in combination with spectral-fitting the lags will provide a powerful tool to constrain black hole mass and spin and study differences in inner region structure between different classes of bright AGN. 

\subsection{Constraining the geometry of the innermost flow and outflow in XRBs and AGN}
\label{ssect:xrb coronae}
 {\it eXTP} provides an entirely novel way to track the causal connection between inflow of mass on to the black hole, and subsequent outflow through a jet. X-rays are expected to be produced both from inflowing material close to the black hole, and also from outflowing material at the base of a jet. Since the spectrum from these two regions is likely to be similar, with both regions producing broad continuum spectra, it is difficult to disentangle their emission spectrally (e.g.~\cite{ref76}). The causal connection between accretion flow and jet can be probed using variability analysis, since fluctuations originating in the flow can propagate up the jet after some delay time.  
For example, in BH XRBs a time lag can be detected between X-rays, emitted mainly in the inflow or `corona' but also partially from the jet base, and infrared emission, thought to be emitted either through cyclo-synchotron processes in the jet \cite{ref77} or - at least in some cases - by synchrotron emission from the hot flow \cite{poutanen&veledina14,veledinaetal13}. This is not currently possible for the case of the jet base however, since both inflowing and outflowing components radiate in X-rays.
\begin{figure*}[t]
\centering
\includegraphics[width=0.47\textwidth]{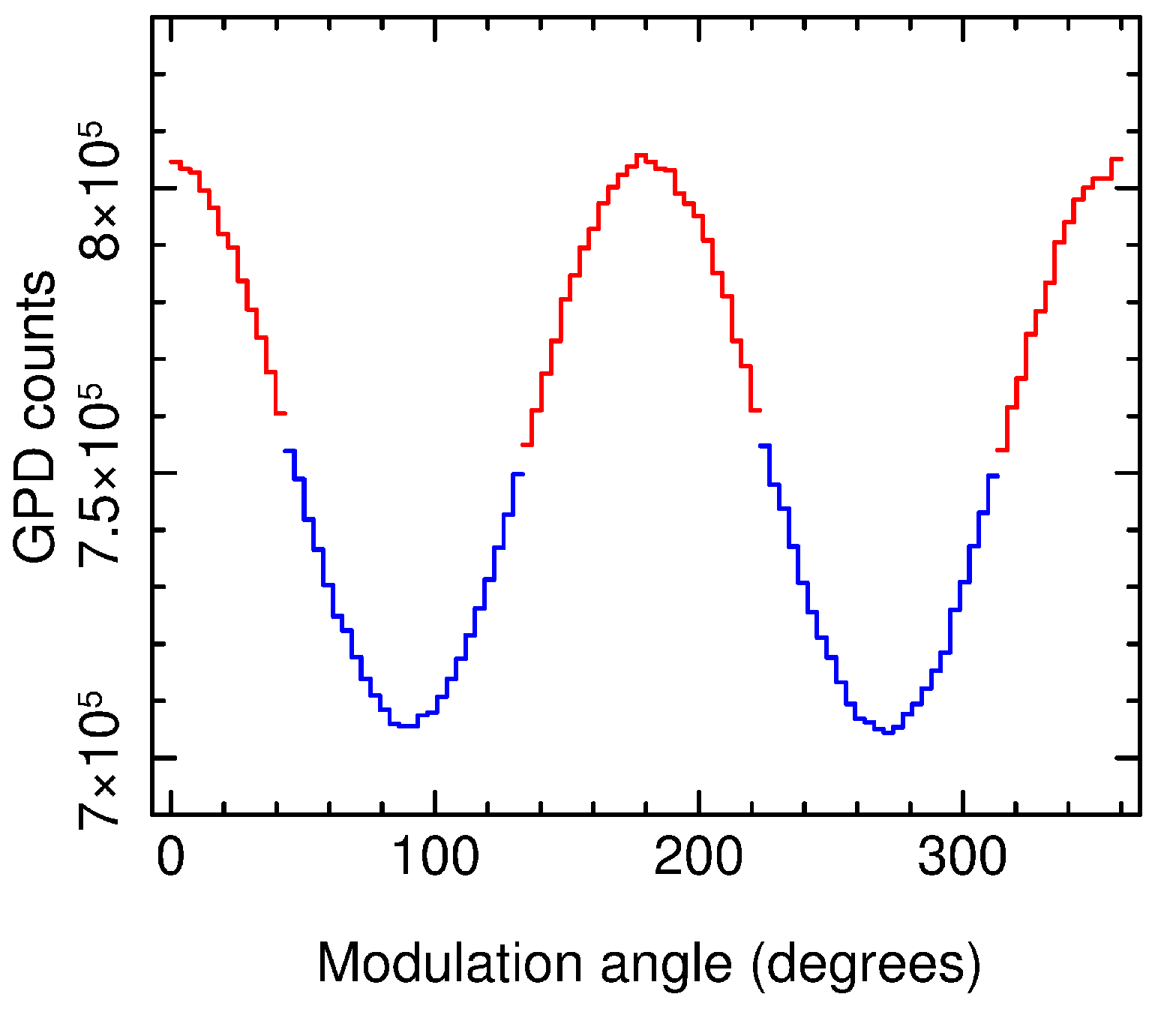}
\includegraphics[width=0.47\textwidth]{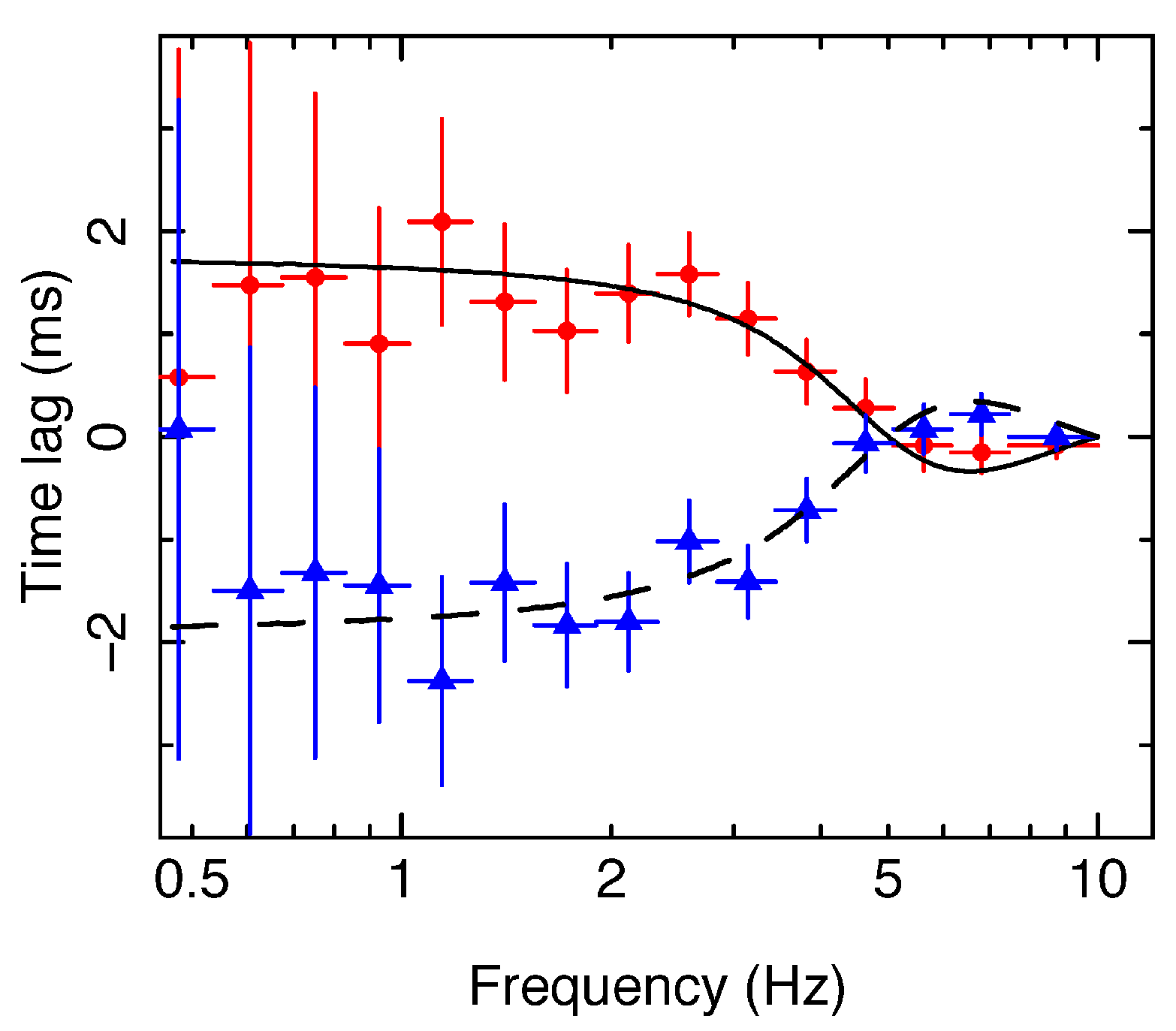}
\caption{
\textit{Left:} PFA counts as a function of modulation angle for a simulated 50 ks observation consisting of a weakly polarized component from the corona and a strongly polarized component from the jet (see text for details). We define high polarization (red) and low polarization (blue) regions. \textit{Right:} Time lag as a function of frequency. Here, high polarization photons \textit{lag} the LAD (red circles) and low polarization photons \textit{lead} the LAD. The black lines (dashed line: low polarization) show the input model, i.e. without Poisson statistics being included in the light curves.
}\label{fig:bbnresults}
\end{figure*}
However, the two emission components could be disentangled using X-ray polarization measurements.
The coronal X-ray emission is expected to be only weakly polarized, whereas photons produced by Compton scattering or synchrotron emission from the non-thermal electrons in a magnetized jet are expected to be strongly polarized. With {\it eXTP}, even if the corona and jet emit identical spectra, with the same polarization angle, we will still be able to detect a time lag resulting from the propagation time between corona and jet. To demonstrate this, we simulate broadband variability originating from the corona, assuming a zero-centered Lorentzian power spectrum with a width of $1$ Hz and a total rms variability amplitude of $20 \%$. We generate the jet light curve by applying a lag of $\tau=0.1$ s to the corona light curve. We assume that the spectra emitted from the jet and corona are identical in shape and also that the polarization angle is the same ($\psi_0=0$) for both regions. We assume that $80\%$ of the total X-ray flux comes from the corona, with the remaining $20\%$ from the jet. The jet has a high polarization degree of $70\%$, whereas the corona has a polarization degree of only $10\%$. 
We simulate {\it eXTP} observing such a system for $50$ ks, assuming a flux and spectrum typical of GRS 1915+105. ~\cref{fig:bbnresults} (left) shows the total counts detected by the SFA as a function of modulation angle (i.e. the direction of the electron track caused by each photon as it hits the GPD detector). We clearly see the sinusoidal pattern expected for a polarized signal, peaking at $\psi_0=0$ degrees. We split this plot up into `high polarization' regions close to the peaks (red) and `low polarization' regions elsewhere (blue). Since the jet emission is much more polarized than the corona emission, the high polarization regions here contain a greater fraction of jet photons than the low polarization regions. Since the jet lags the corona, a light curve of `high polarization' photons should therefore lag a light curve of `low polarization' photons. The lag of $0.1$ s will be reduced by dilution however, since the high polarization region does not contain exclusively jet photons. 

We create high and low polarization light curves by selecting photons only from these red and blue regions. We could calculate the cross-spectrum between these two light curves in order to measure a lag. However, since the effective area of the LAD instrument is much larger than that of the PFA, we can significantly increase signal to noise (by a factor $>7$) by correlating both polarization light curves with the LAD light curve, which contains far more counts. We therefore calculate the cross-spectrum between the high polarization light curve and the LAD data, and also the cross-spectrum between the low polarization light curve and the LAD data and plot the two resulting lag-frequency spectra in~\cref{fig:bbnresults}.  The high polarization light curve does indeed lag the LAD light curve, which contains all jet and corona photons, and the low polarization light curve \textit{leads} the LAD light curve, as expected.  The difference in lag can then be weighted by the polarization degree (which gives the approximate fraction of jet photons) to yield the intrinsic lag of the jet emission relative to the coronal emission.  Using the SFA alone, the errors on the lags would be much larger and thus the lags would not be detectable.  Therefore, this simulation demonstrates that {\it eXTP}'s combination of polarimetry and high-throughput timing-capability can probe the causal connection between corona and jet, even if the two have identical spectral shape.

In radio quiet AGN, where the jet is probably absent, the coronal emission is expected to be polarized, with the polarization percentage depending mainly on the geometry and optical depth of the corona \cite{ref74}.
A  slab-like and a sphere-like disk-corona geometry produce quite similar spectral shapes in the X-ray band, while the polarization is always higher in the slab-like scenario. Time-averaged polarimetric measurements will break the spectral degeneracy for a sizeable sample of bright unobscured AGN. In practice, with 200~ks exposure we will be able to reach a Minimum Detectable Polarization (MDP) of about 2 per cent in a mCrab AGN, which is sufficient to break the geometry's degeneracy.

\section{Black hole astrophysics}
\label{sect:astro}

It is now recognized  that the radiative, as well as the kinetic, output of accreting black holes can influence their surroundings.  The close feedback between the formation and evolution of galaxies and of their central supermassive black hole \cite{dimatteoetal05} involves a variety of physical phenomena. A currently highly debated example is the plausible key role played by the uncollimated winds from AGN in setting the rate at which galaxies evolve \cite{ref78}. 
In order to build a self-consistent and comprehensive picture we need a deep understanding of black hole accretion througth the interplay between accretion and ouflowing components in both AGN and XRBs. 

As widely discussed above,  the X-ray spectra of black holes are rich in emitting/absorbing features which vary on a wide range of time-scales.  In addition to answering fundamental questions related to matter flows under strong field gravity conditions (see Sect. \ref{sect:SFG}), \textit{eXTP} will also play a significant role in answering important astrophysical questions related to black holes.  {\it eXTP} will tackle the many open questions from multiple, complementary directions.  In particular we describe below the main advantages that \textit{eXTP} will offer by studying the surrounding emitting and absorbing gas of accreting black holes in AGN and XRBs.

\begin{figure*}[t]
\centering
\includegraphics[scale=1.4]{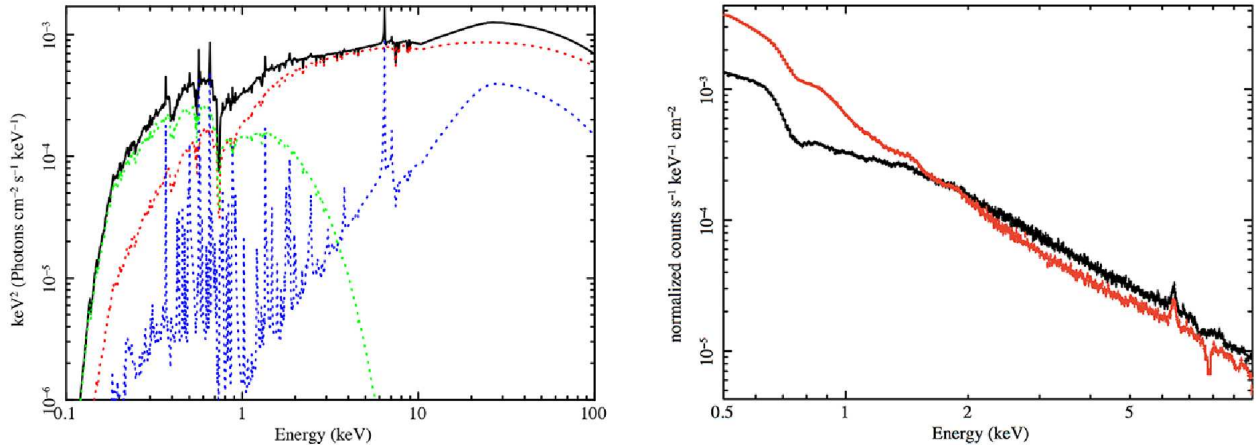}
\caption{\textit{Left panel}: Typical broadband AGN spectral model (black line) used for {\it eXTP} simulations. The several emission/absorption components are shown in different colors: Green: soft X-ray comptonized component; Blue: cold reflection component; Red primary absorbed power-law emission. Warm and ionized outflows are also included in this model. \textit{Right panel}: Simulated {\it eXTP} spectrum from a 100 ks observation of a typical radio-quiet AGN with 2-10 keV flux of 2.5$\times$10$^{-12}$ erg cm$^{-2}$ s$^{-1}$. For the black and red points two different warm absorbers ($\log N_{\rm H}$=22, and 21.5) and UFOs (v=0.1c and 0.15c) components have been considered (see text for details).}\label{fig:astro_AGN}
\end{figure*}

\subsection{The rich environment of supermassive black-holes in AGN}
\label{ssect:astro_agn}

As discussed in Sect. \ref{sect:section1}, the X-ray behaviour of radio-quiet AGN is quite complex and rich with emitting and absorption spectral features that must all be taken into account when modelling the broadband energy spectrum.  Often, multiple absorbing components are present, associated with disk winds (in the form of an outflowing gas) and other circumnuclear regions.  
While the geometry and composition of the outflows can be revealed thanks to their polarimetric signatures \cite{goosmann&matt11} the evolution of the accretion flow associated with changes in the outflow can be probed with great precision in spectroscopy, timing and polarization.

In the environment of AGN there is now strong evidence for at least three absorption components on very different scales: on scales of hundreds of parsecs, on the parsec scale, and within the dust sublimation radius, on sub-parsec scale \cite{ref84,ref85}. The most effective way to estimate the distance of the different absorbers is by means of the analysis of the variability of their column density along the line of sight ($N_{\rm H}$).
In particular, rapid (from a few hours to a few days) variability of the absorbing column, in the form of X-ray eclipses, has been observed in most bright AGN in the local Universe \cite{ref86}, suggesting that obscuration in X-rays is due, at least in part, to BLR clouds. 
Moreover, during the successive covering and uncovering of the inner part of the accretion flow by X-ray eclipsing, a variation of  the polarization degree by a few percent and significant variation (above 10$^\circ$) of the polarization angle are expected  \cite{ref87}. 
Thanks to the PFA MDP  of a few per cent above 1 mCrab, such measurements will be performed in a dedicated sample of bright AGN known to show eclipses \cite{ref88}.

However, most of the astrophysical diagnostics available with \textit{eXTP} will be applicable to the weakest AGN with typical X-ray flux of 10$^{-.12}$ erg cm$^{-2}$s$^{-1}$.
To demonstrate the power of {\it eXTP} for disentangling different components in the spectra of AGN with only moderate flux, we show in~\cref{fig:astro_AGN} the broadband model (left panel) and spectra (right panel) of 100 ks {\it eXTP} observations of typical radio-quiet AGN with 2--10 keV flux $\sim 2.5 \times$10$^{-12}$ erg cm$^{-2}$s$^{-1}$ (relativistic signatures are not included in these simulations, see Sect. \ref{fig:BH_spin}).   The spectra include, in addition to the primary hot Comptonization continuum (produced in the hot corona), a warm ionized absorbing gas component (black points: $\log N_{\rm H}$=22, red points: $\log N_{\rm H}$=21.5) an ultra-fast ionized outflow (black points $v/c$=0.1 and red points: $v/c$=0.15) and a cold distant reflecting medium.  From the {\it eXTP} spectra we will recover key parameters of the primary continuum and the reprocessed/absorbed components with outstanding statistical precision. We remark that more than 400 AGN with this flux level or larger are expected from the well-known $\log N$--$\log S$ distributions \cite{ref89}, thus {\it eXTP} will allow us to build large samples with different selection criteria and X-ray spectra at CCD-quality resolution with extremely good S/N.
\begin{figure*}[ht]
\centering
\includegraphics[scale=0.9, angle=90]{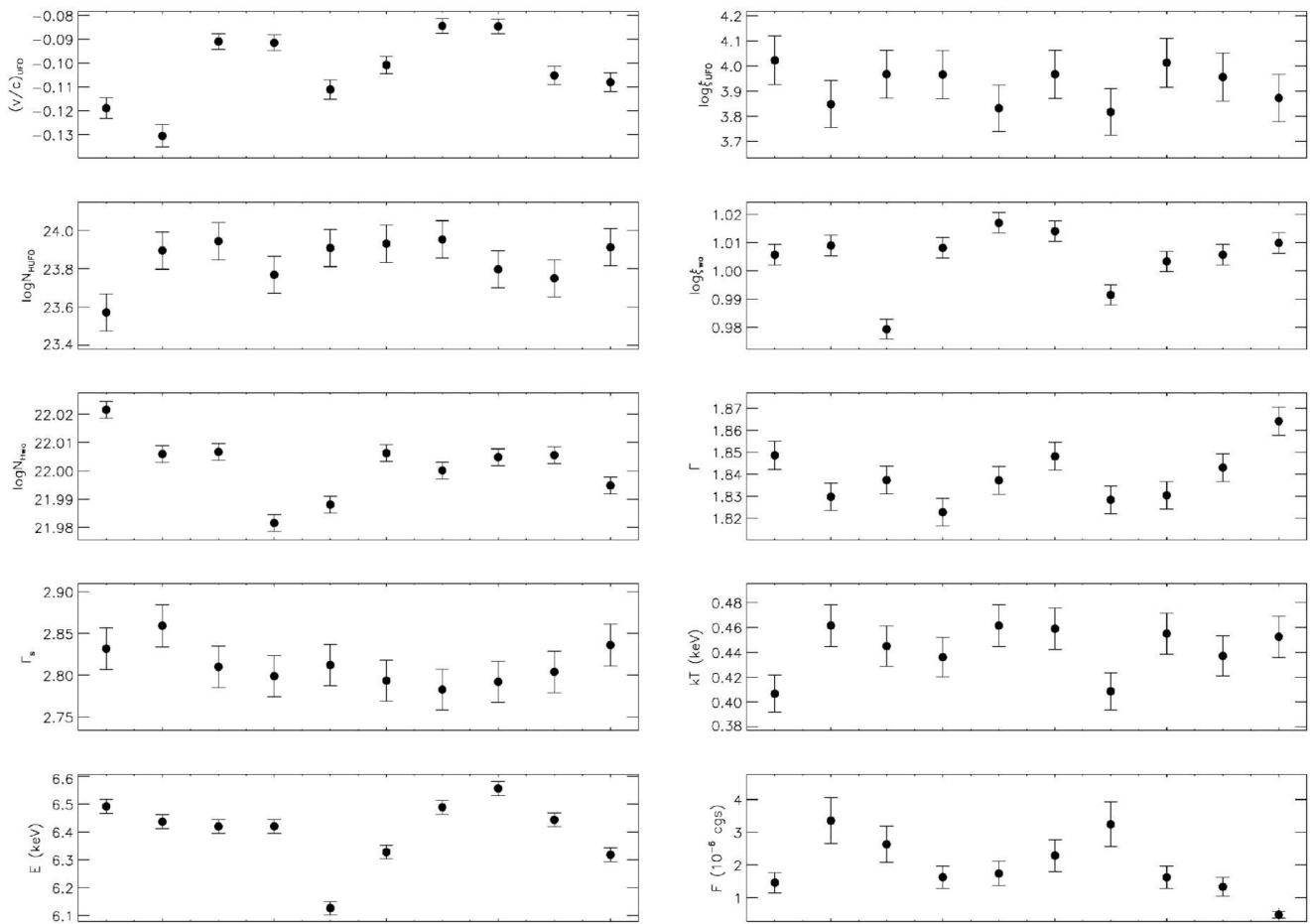}
\caption{Best fit values as obtained with 10 shots of 10~ks exposure observations of a weak AGN of flux 2.5$\times$10$^{-12}$ erg cm$^{-2}$s$^{-1}$. {\it eXTP} will allow us to recover with great precision the primary emission and the physical and geometrical properties of the circumnuclear components, namely: velocity, column density and ionization parameter for both the fast outflows (UFO) and the warm absorber gases ($N_{\rm H}$, $v/c$ and $\log \xi$); primary emission ($\Gamma$), soft X-ray Comptonized component ($\Gamma_s$, $kT$), narrow Fe line (energy and flux).}
\label{fig:astro_AGN2}
\end{figure*}

The large effective area available with {\it eXTP} will also allow for a complementary approach using variability. In addition to using spectra with long integrated exposures, multiple shorter exposures of the same object will allow for variability studies of key parameters of the primary and reprocessed components on relevant time-scales (from tens of ks to years). For example, we show in~\cref{fig:astro_AGN2} the precision we will have for short (10 ks) exposures of the same AGN simulated in~\cref{fig:astro_AGN}. The possibility to follow the temporal behaviour of all these parameters for a large number of objects will open new avenues in the investigation of the properties of the circumnuclear matter in AGN.
Time resolved spectroscopy will allow us to address the systematic error due to models degeneracy when time average tecniques are used.

Another aspect of timing technique is rapresented by the variability studies on long time scale (e.g. with multiple observations on the same target or monitoring programs), which will  provide important contraints on the so-called Unified model for AGN \cite{antonucci93}.
These models invoke the presence of a large-scale molecular ``torus" (i.e. an axisymmetric, circumnuclear absorber), but although there is substantial evidence for such circumnuclear gas, its physical and geometrical structure is quite complex and largely unknown.  X-ray spectral polarimetric measurements are crucial in order to investigate this issue, since the geometry and composition of the torus will directly impact the amount of polarized flux. The degree and position angle of polarization are expected to vary on different timescales compared to the inner regions, leading to different variability profiles for repeated {\it eXTP} observations \cite{marinetal16}. 
The 2-- 10 keV polarization is thus expected to range from a few per cent  for unobscured sources (i.e. for type 1 objects seen almost face-on, as envisaged by the unification  model) up to several tens of per cent for obscured objects (i.e. the type 2 AGN seen edge-on by the observer) \cite{goosmann&matt11}.
With $\sim 100$~ks exposure {\it eXTP} will be able to perform such polarization measurements for the brightest obscured AGN (flux above 10$^{-11}$ erg cm $^{-2}$ s$^{-1}$) with an MDP of 4 per cent.

\subsection{Disk winds in X-ray binaries}
\label{ssect:astro_xrb}

Accretion disk winds have received a great deal of attention in the last few years, thanks to a number of results, both in stellar-mass and in supermassive black holes \cite{ref79,ref90,ref91}. Nevertheless, there are still many unknowns, which need to be studied and resolved, including: the launching mechanism(s) and region, the wind power and structure, the outflowing mass rate, the relation with their collimated ``cousins'' (the jets) and the trigger of their appearance during the source evolution.  Such winds can be extremely variable. When studying stellar-mass black holes, the relevant physical quantities can vary on very short time scales, sometimes even of the order of a few seconds. Thus, time-resolved spectroscopy is the crucial way forward to tackle the above unknowns.

\begin{figure*}[ht!]
\centering
\includegraphics[scale=1.0]{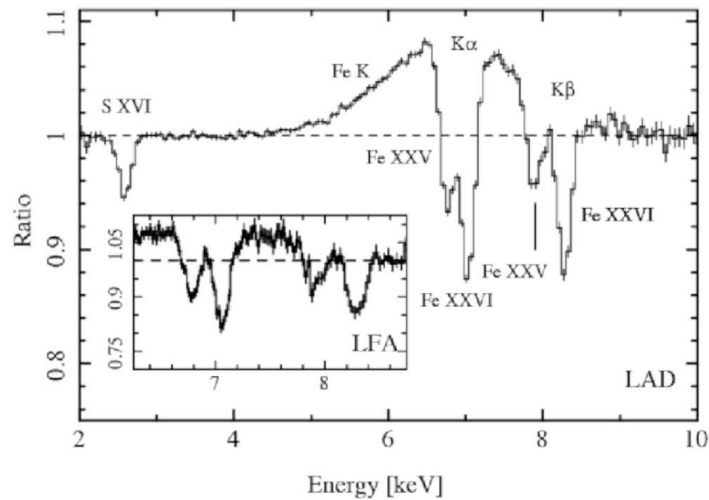}
\caption{Simulated 1 ks LAD exposure of 4U 1630-47 (0.3 Crab, 2-10 keV absorbed flux, continuum and line parameters based on 
\cite{DiazTrigoetal2014}). The Fe XXV and Fe XXVI lines are clearly resolved into K$\alpha$ and K$\beta$ components.
The great precision of the line measurements will allow monitoring of the evolution of such winds over a large range of accretion rates.
}\label{fig:xrb}
\end{figure*}

Such winds are usually identified via their typical marker: a strong (equivalent width of $\sim$35--40 \,eV) Fe~\textsc{xxvi} absorption line. For a typical source flux of 0.2 Crab, {\it eXTP} will detect such strong lines at $3\sigma$ confidence in a little over 7 seconds, a few hundred times faster than Chandra. For brighter sources, around 1 Crab (for example in the case of a nearby binary, or in the very luminous states observed in some sources), a similar detection will be achieved in as little as 1--2 seconds, with a huge improvement on variability studies. These outstanding performances will allow a new type of study: the measure of the turbulence in the wind structure via the absorption line variability, when compared with the observed X-ray luminosity. Simulations show that, for typical wind parameters ($N_{\rm H}=5\times10^{22}\,\mathrm{cm}^{-2}$ and logarithm of ionisation parameter, $\log\xi=4.25$) the effects of an X-ray luminosity variation by as little as 10\% on the ionisation parameter could be detected at well above 3$\sigma$ (reducing to a $\sim 97\%$ confidence for a 6\% luminosity variation), allowing any further (anomalous or even uncorrelated) ionisation variability to be used to map the wind density structure, by disentangling the effects of photo-ionisation from the clumpiness of the wind.
Once again, the combination of focussing X-ray telescopes and LAD detectors will allow these lines to be tracked during the launching and quenching of the winds down to  source at low luminosities. Namely, {\it eXTP} will detect weaker lines (7.5 eV equivalent width) at 3$\sigma$ confidence in less than 200 seconds for a 0.2 Crab source and in 1 ks for 0.03--0.04 Crab. Even for a source as faint as 10 mCrab, {\it eXTP} will need less than an hour to detect such a weak line (see ~Figure \ref{fig:xrb}). 
Finally, the good spectral resolution of the SFA will allow detection of line velocities as low as 300 km~s$^{-1}$, at 5$\sigma$ significance, in as little as 100 seconds.  All in all, {\it eXTP} will enable a leap forward in the study of accretion disk winds in stellar-mass black holes, allowing monitoring of the evolution of such winds over a large range of accretion rates, and pinpointing the epochs of their appearance/disappearance.

\section{{\it eXTP} science in the gravitational wave astronomy era}
\label{sect:sect_GW}

The detections by the LIGO-Virgo experiment \cite{ref92,abbottetal17c} of gravitational radiation from merging black holes and neutron stars mark the dawn of the era of gravitational wave astronomy. Gravitational wave detectors are a new, powerful tool to study the strong-field regime of gravity and to probe physics beyond the standard model. In this context, {\it eXTP} can be extremely useful to better understand the strong-gravity environment of black holes and neutron stars, and it is
complementary to gravitational wave detectors to test gravity and fundamental physics under extreme conditions.
Gravitational wave interferometers target the inspiral, merger, and ringdown of compact objects, where spacetime is being shaken by closely orbiting masses and hence is dynamic~\cite{sathyaprakashetal2009}. In contrast, {\it eXTP} probes the stationary spacetime metric of compact objects through the X-ray emitting plasma of the accretion disk, essentially a luminous test fluid orbiting the black hole with negligible self-gravity.
\textit{eXTP} uniquely covers not only supermassive but also stellar-mass black holes, and does so in completely analogous settings, performing comparative studies of black holes spanning a factor 10$^{8}$ in mass via well-established diagnostics, using observables already clearly identified in current observations. Spacetime curvatures are small near SMBH event horizons but scale as 1/BH mass$^{2}$, and hence are a factor ~10$^{16}$ higher near stellar-mass black holes. X-ray observations uniquely provide access to these very-high curvature stationary metrics. GR predicts orbital dynamics are not affected by curvature, and accretion flows across the black-hole mass scale probe this prediction over 16 orders of magnitude. 

The X-ray diagnostics discussed in this paper provide a unique test to confirm GR theory into the strong field regime, by confirming that no systematic departures are found from GR predictions of the motion of light and matter across 8 orders of magnitude of black hole mass.  Conversely, if modification of dynamics is required over and above all of the effects that might alter purely geodetic motion (due to the complex astrophysical environment close to black holes, see Sect. \ref{sect:astro}), this would violate the no-hair theorem and new theories will need to be explored.

In this section we report the most compelling examples of the information that {\it eXTP} will provide in the context of the gravitational waves astrophysics. In particular, we outline below the {\it eXTP}  key role in the multi-messanger astrophysical domain, the fundamental physics study and the standard model for particles physics.

\subsection{Detecting electromagnetic counterparts of gravitational wave events}

\textit{eXTP} has unique capabilities for the detection and study of electromagnetic signals as the counterparts associated with many sources of gravitational waves.
The WFM, with its large field of view and sensitivity down to the soft energies, offers the potential to locate mergers of compact binary systems in the sky through the detection of an electromagnetic counterpart, as recently observed in the landmark multimessenger observation of the binary neutron star coalescence GW170817 ~\cite{abbottetal17c} producing a kilonova, the short-and GRB 170817A ~\cite{abbottetal17b}. Sky localization improves our understanding of the gravitational wave data, because the parameters of the merging system can be determined much more accurately if the location of the source is known. 
Almost all short GRBs are accompanied by an X-ray afterglow, and the recent observations indicate that in a large fraction of events a long-lived neutron star may be formed rather than a black hole.
LAD re-pointing in response to gravitational wave triggers will allow the X-ray features of the afterglow to be sytudied and characterized with great precision. Unambiguous electromagnetic signatures of the post merger event will allow us to address the open questions concerning the nature of short GRB central engines and put important constraints on models \cite{siegel&ciolfi16a,siegel&ciolfi16b}.

Moreover, a knowledge of the redshift of the source can be used to measure the Hubble constant \cite{ref93,abbottetal17a} or to compare the propagation properties of electromagnetic and gravitational waves, thus
constraining (say) Lorentz-violating theories, that predict a different speed for electromagnetic and gravitational waves~\cite{ref94,abbottetal17b}.

\subsection{Testing general relativity and the nature of black holes}

General relativity has passed weak-field, binary-pulsar \cite{kramer16}, and even the most recent strong-field tests provided by the observation of black hole mergers with flying colors~\cite{ref95,ref94,ref96}. Nevertheless, relativistic
theories that modify general relativity only at the scales of compact objects are still compatible with
observations~\cite{ref96}. For instance, quadratic gravity theories (such as Einstein-dilaton-Gauss-Bonnet and Chern-Simons gravity) differ from general relativity only in large-curvature regions, such as near the horizon of stellar-mass black holes~\cite{ref94}.
Current and future observations will be instrumental to test general relativity in the strong-field
regime, and to rule out or detect exotic compact objects which can compete with or coexist with black
holes and neutron stars~\cite{ref97}. 

One of the most subtle consequences of general relativity is the so-called ``no-hair'' theorem, stating that stationary black holes can be fully characterized by their mass, spin and charge \cite{ref99}.  
If general relativity is modified in the strong-field regime, or if black holes are replaced by more exotic compact objects, the no-hair theorem is generally violated. 
As discussed in Sect.~\ref{sect:SFG}, the two common approaches to test the no-hair theorem from the X-ray emission of accreting black holes are the profile of the iron $K\alpha$ line~\cite{fabianetal1989} and the intensity of the thermal component of the black hole disk~\cite{ref100}. 
These approaches allow the measurement of the location of the ISCO, from which it may be possible to extract not only the mass and spin of the black hole, but also its quadrupole moment (which depend only on the black hole spin and mass because of the \textit{no-hair} theorem, see e.g. ref.~\cite{ref101,ref102}), providing constraints on the departure of space-time from the Kerr metric.

Nevertheless, as widely discussed in the previous sections, the central regions of compact objects consist of multiple variable emission structures and the spectrum may further be subject to complex large scale emission and absorption.  These astrophysical effects should be correctly modeled when looking for tiny effects possible due to departures from modified GR.  eXTP is designed to be a very sensitive probe of these emitting regions by using multiple, independent approaches to determine the geometry and dynamics of the central regions and disentangle the effects of the wider environment.  By accounting for the astrophysical effects, eXTP is therefore one of the most promising experimental tools to constrain violations of the no-hair theorem in the near future, by measuring the mass, spin and quadrupole of black holes and enabling the detection of possible weak departures from the non GR metric.

Another promising diagnostic is the measurement of QPOs in the X-ray emission from accreting black
holes (see Sect.~\ref{ssect:qpo} and ref.~\cite{ref103}). For instance, the detection of two QPO triplets (a pair of high frequency QPOs and a low frequency QPO) by an X-ray instrument with the large
effective area provided by the LAD would give stringent constraints on modified gravity theories such as
Einstein-dilaton-Gauss-Bonnet gravity, as long as the black hole spin is $a\gtrsim0.5$~\cite{ref104}.

\subsection{Black-hole spin measurements and searches for fundamental fields beyond the standard model}

In recent years it was realized that astrophysical black-hole observations have surprising implications for dark matter searches, allowing us to constrain or reveal new physics beyond the Standard Model. In particular, stellar and massive black holes are unique laboratories to search for ultralight bosonic fields such as QCD axions, dark photons, or those
emerging in fuzzy dark-matter models (for two recent reviews in the context of astrophysics and cosmology see ~\cite{ref105} and \cite{marsh16}). The reason is that ultralight bosonic fields around spinning black holes can trigger superradiant instabilities, forming a long-lived condensate outside the horizon~\cite{ref106}. 

Superradiant instabilities spin black holes down and they can dramatically affect the dynamics of astrophysical black holes, providing a portal for astrophysical tests of bosonic dark matter in the poorly
explored range below $10^{-10}\,{\rm eV}$~\cite{ref107}.
The most striking consequence of this scenario is that observations of fast-spinning black holes would be disfavored if ultralight bosons in a certain mass range exist in nature. Thus, black-hole mass and spin measurements can be translated
into bounds on (or indirect evidence for) ultralight fields.  Given the wide range of boson masses still unconstrained
to date, it is crucial to gather a statistically significant number of mass and spin measurements, both for stellar and
for supermassive black holes (and possibly also for intermediate-mass black-hole candidates).

In this context, {\it eXTP} will allow for measurements of the spin of accreting black holes with great precision (which is comparable or better than current gravitational-wave detectors at design sensitivity) and accuracy [cf.~\cref{fig:BH_spin}], thus allowing  to constrain the masses of ultralight bosons.

Similar synergies between astrophysical observations and fundamental physics have recently been explored in the context of gravitational-wave astronomy~\cite{ref108}, but their connection to future X-ray observations deserves further study. Constraints become more stringent as the number of observations increases, and therefore the large set of black-hole mass and spin measurements provided by future gravitational-wave and X-ray detectors will allow us to constrain the properties of dark matter in a wide range of masses between $10^{-19}\,{\rm eV}$ and $10^{-11}\,{\rm eV}$.

\section{Summary}
\label{summary}

We have described in this paper the unprecedented progress that \textit{eXTP} will make in the study of matter accreting in the strong field gravity regime in the near-environment of black holes.
Thanks to the combination of the instruments on board, \textit{eXTP} will measure the same physical quantity using different diagnostics. 
In particular, the large {\it eXTP} area will allow us to greatly improve the statistical photon counting errors, and simultaneously to address the systematic error generated from the degeneracy involved with time-averaged spectral modelling. Observing how the spectrum evolves with time or with QPO phase (for XRBs) provides many more independent data points than the time-averaged spectrum, with different time bin or phase preferentially sampling the relativistic distortions of different parts of the accretion disk.

We summarize below  the three independent diagnostics provided by {\it eXTP} data, in order to investigate the behaviour of matter in the strong-field gravity regime  and the geometry and accretion physics in the innermost region around accreting black holes.  
\begin{itemize}

\item[-] \textit{Relativistically broadened disk reflection.} The disk reflection components (namely the iron line and the Compton bump) will be measured with great precision by {\it eXTP} in spectra of XRBs and AGN (see Sect.~\ref{ssect:reflection}) using the state-of-the-art models to measure black hole spin. In XRBs the virtually pile-up free LAD data, will allow us to measure changes in the accretion flow around the black hole on unprecedentedly short time-scales.  In AGN Doppler tomography measurements (see Sect.~\ref{ssect:AGNtomo}) will break  model degeneracies due to possible complex absorption components (alternative to the relativistic models). In fact, since these measurements are based on variability in the residuals in the iron region, any contribution from a narrow line in the profile (arising by necessity at larger radii and hence varying much slower than the 10 ks inner disk orbital period) will drop out automatically.

\item[-] \textit{Continuum fitting.} 
By combining spectral and polarimetric diagnostics, the use of disk thermal emission to map the innermost regions and measure black hole spin will greatly improve in accuracy (see sect. \ref{ssect:continuum_fitting}).
The WFM  will select the soft states (dominated by the disk thermal component with minimal  contamination by power-law emission) best suited for spin measurements. The disk parameters will be determined thanks to the SFA and LAD broad energy coverage, with absorption features in the soft state accurately modelled (see Sect.~\ref{ssect:astro_xrb}).
Measurement of the rotation with energy of the polarization angle of the thermal emission in XRBs will provide a futher independent method to measure the black hole spin.  As a third consistency check, the estimated spin can be compared with that obtained from fitting the relativistically broadened disk reflection, which should be significantly more accurate than current measures (see Sect.~\ref{ssect:reflection}). 

\item[-] \textit{QPOs.} Another independent measurement of black hole  mass and spin will be 
provided by modelling the low and high-frequency QPOs observed in the intermediate states prior to or following the soft state in XRBs (see Sect.~\ref{ssect:qpo}). Low-frequency QPOs seen in the hard and hard-intermediate states may be produced by Lense-Thirring precession of the inner hot flow \cite{ref62}.  In this case, modelling the geometry of the precessing flow using tomography and timing-polarimetry (see Sect.~\ref{ssect:QPOtomo} and \ref{ssect:xrb coronae}), will give a direct measure of the offset between the black hole spin and disk angular momentum, allowing us to test the effects of such an offset on the disk inner radius measured in the soft state, when it extends to the ISCO. 

\end{itemize}

We have also shown that {\it eXTP}  will address many fundamental issues in the wider context related to astrophysics of accreting black holes: the structure of the innermost regions as they evolve through different accretion states, the associated production of jets and winds and their link with AGN feedback and hence galaxy evolution:  
\begin{itemize}

\item[-] \textit{The innermost regions around accreting black holes.}
Reverberation lags for the broad Fe K emission and reflection hump in XRBs and AGN (see Sect.~\ref{ssect:reverberation}), allow a further independent measure of the geometry of the innermost part of the accretion flow (in particular the disc-corona regions) down to few r$_g$ scale.
{\it eXTP} will track the causal connection between inflow of mass on to the black hole, and subsequent outflow through a jet (see Sect.~\ref{ssect:xrb coronae}).  The spectrum from these two regions is likely similar, but the causal connection between accretion flow and jet can be probed using variability analysis, since fluctuations originating in the flow can propagate up the jet after some delay time.  The corona is expected to be only weakly polarized, whereas the jet is expected to be strongly polarized. Even if the corona and jet emit identical spectra, with the same polarization angle, we will still be able to detect a time lag resulting from the propagation time between corona and jet.  

\item[-] \textit{The complex environment of AGN.}
Time resolved spectral study of AGN (see Sect.~\ref{ssect:astro_agn}) will allow us to measure the variability of the key parameters of the primary and reprocessed components (ionized and cold absorbers, distant reflection, fast outflows) on relevant time-scales (from tens of ks to years).  The possibility to follow the temporal behaviour of all spectral parameters for a large number of objects will open new avenues in the investigation of the physical properties and geometry of the circumnuclear matter in AGN.
Moreover, time average polarimetric measurements avilable with \textit{eXTP} will be crucial in order to investigate the geometry of the molecular `torus' and the corona invoked in AGN unification models.

\item[-] \textit{Disk winds in XRBs.}
The combination of SFA and LAD detectors will allow us to track the rapidly-varying absorbtion lines in XRBs (see Sect.~\ref{ssect:astro_xrb}) during the launching and quenching of the winds, down to the low luminosities.
This capability will open up a new type of study: the measurement of the turbulence and density structure in the wind via absorption line variability, when compared with the observed X-ray luminosity.

\end{itemize}

 Within the context of the new gravitational wave astrophysics,  and in a fully complementary view with respect to gravitational wave interferometers, \textit{eXTP} will address the stationary spacetime metrics of compact objects (see Sect.~\ref{sect:sect_GW}), probed by a luminous test fluid orbiting the black hole with completely negligible self-gravity.
\textit{eXTP} uniquely will perform comparative studies of black holes spanning a factor 10$^{8}$ in mass (thus ~10$^{16}$ in curvature) via well-established diagnostics. Accretion flows across the black-hole mass scale will allow us to probe the GR predictions over 16 orders of magnitude. 

\Acknowledgements{
The authors would like to thank the referee for the constructive comments and suggestions. We also thank the external referees for their advise prior to submission.
The Italian collaboration acknowledges financial contribution from the agreement ASI-INAF n.2017-14-H.O. The Chinese team acknowledges the support of the Chinese Academy of Sciences through  the Strategic Priority Research Program of the Chinese Academy of Sciences, Grant No. XDA15020100. A. Zdziarski acknowledges the Polish National Science Centre grant 2013/10/M/ST9/00729.
}


{\bf{Author Contributions}} This paper is an initiative of eXTP’s Science Working Groups 2 on  Strong Field Gravity and 5 on  synergy with Gravitational Waves, whose members are representatives of the astronomical community at large with a scientific interest in pursuing the successful implementation of \textit{eXTP}. The paper was primarily
written by Alessandra De Rosa, Phil Uttley, Lijun Gou, Yuan Liu, with major contributions by 
Cosimo Bambi, Emanuele Berti, Marco Feroci, Valeria Ferrari,  Leonardo Gualtieri, Paolo Pani (Gravitational Waves);
Didier Barret, Tomaso Belloni,  Adam Ingram, Sara Motta (QPOs and QPO phase resolved spectral polarimetry)
Giorgio Matt,   Vladimir Karas, Ilaria Caiazzo, Jeremy Heyl (X-ray polarimetry);  Stefano Bianchi, Piergiorgio Casella, Bin Luo, Joseph Neilsen,  Xinwen Shu, Junfeng Wang, Jian-Min Wang, Yongquan Xue, Weimin Yuan, Yefei Yuan (AGN and XRB astrophysics).
Contributions were edited by Alessandra De Rosa and Phil Uttley. Other co-authors
provided input to refine the paper.











\end{multicols}

\begin{thebibliography}{99}
\label{references}

\bibitem {volonterietal2005}   M.Volonteri, P. Madau, E. Quataert, M.J. Rees, Astrophys. J. \textbf{620}, 69 (2005)

\bibitem {sathyaprakashetal2009}   Sathyaprakash B.S., Schutz B.F., 2009, Living Reviews in Relativity 12 (2009)

\bibitem {fabianetal1989}    A. C. Fabian, M. Rees, L. Stella, N.E. White, Mon.Not.R.Astron. Soc. \textbf{238}, 729 (1989).

\bibitem {ref2}   Y. Tanaka, K. Nandra, A.C. Fabian, et al., Nature \textbf{375}, 659 (1995) 

\bibitem {ref17}   J.M. Miller., Annual Rev. Astron. Astrophys. \textbf{45}, 441 (2007)

\bibitem {stellaetal1999}  L. Stella, M. Vietri, S.M. Morsink, Astrophys. J. \textbf{524}, L63 (1999)

\bibitem {ref4}    M. van der Klis, F. Jansen, J. van Paradijs, et al., Nature \textbf{316}, 225 (1985) 

\bibitem {ref5}   S. Miyamoto, K. Kimura, S. Kitamoto ,T. Dotani, K. Ebisawa, Astrophys. J. \textbf{383}, 784 (1991)

\bibitem {ref6}   T.E. Strohmayer, W. Zhang, J.H. Swank, et al., Astrophys. J. \textbf{469}, L9 (1996)

\bibitem {ref7}   M. van der Klis, J.H. Swank, W. Zhang, et al., Astrophys. J. \textbf{469}, L1  (1996)

\bibitem {ref8}    R.A. Remillard, J.E. McClintock, G.J. Sobczak, et al.,  Astrophys. J. \textbf{517}, L127 (1999)

\bibitem {ref9}    M. Gierli\'{n}ski, M. Middleton, M. Ward, C. Done, Nature \textbf{455}, 369 (2008)

\bibitem {ref10}   G. Matt, A.C. Fabian, R.R Ross, Mon.Not.R.Astron. Soc.\textbf{262}, 179  (1993)

\bibitem {ref11}   A. Martocchia, V. Karas, G. Matt, Mon.Not.R.Astron. Soc. \textbf{312}, 817 (2000)

\bibitem {ref12}   K. Nandra, P.M. O'Neill, I.M. George, J.N. Reeves J.N., Mon.Not.R.Astron. Soc. \textbf{382}, 194 (2007)

\bibitem {ref13}   I. de La Calle P{\'e}rez, A.L. Longinotti, M. Guainazzi, et al., Astron. Astrophys. \textbf{524}, A50 (2010)

\bibitem {ref14}   G. Risaliti, F.A. Harrison, K.K. Madsen, et al., Nature \textbf{494}, 449 (2013)

\bibitem {ref15}   A. Marinucci, G. Matt, E. Kara, et al., Mon.Not.R.Astron. Soc. \textbf{440}, 2347 (2014)

\bibitem {ref16}   J.A. Tomsick, M.A. Nowak, M.Parker, et al., Astrophys. J. \textbf{780}, 78 (2014)

\bibitem {ref28}   S.N. Zhang, W. Cui,W. Chen, Astrophys. J. \textbf{482}, L155 (1997)

\bibitem {ref18}   J.E. McClintock, R. Narayan, S.W. Davis, et al., Classical and Quantum Gravity \textbf{28}, 114009 (2011)

\bibitem {vanderKlis2006}  M. van der Klis, Compact stellar X-ray sources.  Cambridge Astrophysics Series, \textbf{39}, 39-112 (2006)

\bibitem {ref62}  L. Stella, M. Vietri, Astrophys. J. \textbf{492}, L59 (1998)

\bibitem {Mottaetal2014a} S. Motta, T.M. Belloni, L. Stella, T. Mu\~{n}oz-Darias, R. Fender, Mon.Not.R.Astron. Soc. \textbf{437}, 2554 (2014)

\bibitem {Mottaetal2014b} S. Motta, T. Mu\~{n}oz-Darias, Sanna, A., R. Fender, T.M. Belloni, L. Stella,  Mon.Not.R.Astron. Soc. \textbf{439}, L65 (2014)

\bibitem {Doneetal2007} C. Done, M. Gierli\'{n}ski, A. Kubota, Astron. Astrophys. R. \textbf{15}, 1 (2007)

\bibitem {ref48}   R. Remillard, J. McClintock, Annual Rev. Astron. Astrophys. \textbf{44}, 49 (2006)

\bibitem {ref49}   T.M. Belloni, A. Sanna, M. M{\'e}ndez, Mon.Not.R.Astron. Soc. \textbf{426}, 1701 (2012)

\bibitem{Motta2016} S.E. Motta, Astronomische Nachrichten, \textbf{337}, 398 (2016)

\bibitem{antonucci93} Antonucci, R.,   Annual Rev. Astron. Astrophys., \textbf{31}, p. 473, (1993)

\bibitem {ref79}   J. Neilsen, J.C. Lee, Nature \textbf{458}, 481 (2009)

\bibitem {ref90}   G. Ponti, R.P. Fender, M.C. Begelman, et al., Mon.Not.R.Astron. Soc. \textbf{422}, 11 (2012)

\bibitem {ref91}  F. Tombesi, M. Cappi, J.N. Reeves, et al., Mon.Not.R.Astron. Soc. \textbf{430}, 1102 (2013)

\bibitem {ref81}   F. Tombesi, M. Cappi, J.N. Reeves, et al., Astron. Astrophys. \textbf{521}, A57 (2010)

\bibitem {dimatteoetal05} T. Di Matteo, V. Springel, L. Hernquist, Nature \textbf{433}, 604 (2005)

\bibitem {connors&stark77} P.A. Connors, R. F. Stark, Nature, \textbf{269}, 128 (1977)

\bibitem {connorsetal80} P. A.Connors, R. F. Stark, T. Piran, Astrophys. J. \textbf{235}, 224 (1980) 

\bibitem {dovciaketal08} M. Dov{\v c}iak, F.  Muleri, R. W. Goosmann, et al., Mon.Not.R.Astron. Soc.  \textbf{391}, 32 (2008)

\bibitem {matt93} G. Matt, A. C. Fabian, R. R., Ross., Mon.Not.R.Astron. Soc.  \textbf{262}, 179 

\bibitem {ref72}   J.D. Schnittman, J.H. Krolik, Astrophys. J. \textbf{701}, 1175 (2009)

\bibitem {ref74}   J.D. Schnittman, J.H. Krolik, Astrophys. J. \textbf{712}, 908 (2010)  

\bibitem {wattsetal18} A.L. Watts, Yu W., J. Poutanen, Zhang S., et al, 2018, Science China, this issue (eXTP White Paper on Dense Matter)

\bibitem {fengetal18} Feng H., A. Santangelo, S. Zane, et al., 2018, Science
China, this issue (eXTP White Paper on Strong Magnetism)

\bibitem {intzandetal18} J. in ’t Zand, E. Bozzo, Qu J., et al., 2018, Science China, this issue (eXTP White Paper on Observatory Science)

\bibitem {zhangetal18} Zhang S.N., A. Santangelo, M. Feroci, et al., 2018, Science China, this issue (eXTP White Paper on Instrumentation)

\bibitem {ref22}   T. Dauser, J. Garcia, J. Wilms, et al., Mon.Not.R.Astron. Soc. \textbf{430}, 1694 (2013)

\bibitem {Garcia2014}  J. Garc\'ia J., T. Dauser, A. Lohfink, T.R. Kallman et al., Astrophys. J. \textbf{782}, 76 (2014)

\bibitem {Garcia2016}  J. Garc\'ia, A.C. Fabian, T.R. Kallman, T. Dauser, M.L. Parker, 
J.E. McClintock, J.F. Steiner, J. Wilms, Mon.Not.R.Astron. Soc. \textbf{462}, 751 (2016)

\bibitem {ref23}   D.R. Wilkins, A.C. Fabian, Mon.Not.R.Astron. Soc. \textbf{430}, 247 (2013)

\bibitem {ref24}   T. Dauser, J. Garcia, M.L. Parker, A.C. Fabian, J. Wilms, Mon.Not.R.Astron. Soc. \textbf{444}, L100 (2014)

\bibitem {ref25}   C.S. Reynolds, Classical and Quantum Gravity \textbf{30}, 244004 (2013)

\bibitem {milleretal09}	Miller, L.; Turner, T. J.; Reeves, J. N., Mon.Not.R.Astron. Soc. \textbf{399}, L69 (2009)

\bibitem {donediaz10}	Done, C.; Diaz Trigo, M., Mon.Not.R.Astron. Soc. \textbf{407}, 2287 (2010)

\bibitem {ref26}   R.C. Reis, A.C. Fabian, R.R. Ross, J.M. Miller, Mon.Not.R.Astron. Soc. \textbf{395}, 1257 (2009)

\bibitem {ref27}    M. D{\'{\i}}az Trigo, A.N. Parmar, J. Miller, E. {Kuulkers}, M.D. Caballero-Garc{\'{\i}}a, Astron. Astrophys. \textbf{462}, 657 (2007)

\bibitem {ref29}   J.E. McClintock, R. Narayan, J.F. Steiner, Space Sci. Rev.\textbf{183}, 295 (2014)

\bibitem {ref30}   J.F. Steiner, J.E. McClintock, R.A. Remillard, L. Gou, S. Yamada, R. Narayan, Astrophys. J. \textbf{718}, L117 (2010)

\bibitem {ref31}   Y. Zhu, S.W. Davis, R. Narayan, et al., Mon.Not.R.Astron. Soc. \textbf{424}, 2504 (2012)

\bibitem {ref35}   M.J. Reid, J.E. McClintock, J.F. Steiner, et al., Astrophys. J. \textbf{796}, 2 (2014)

\bibitem {ref33}   L. Gou, J.E. McClintock, R.A. Remillard, et al., Astrophys. J. \textbf{790}, 29 (2014)
\bibitem {ref39}   Z. Chen, L. Gou, J.E. McClintock, J.F. Steiner, J. Wu, W. Xu, J. Orosz, Y. Xiang, Astrophys. J., \textbf{825}, 45  (2016)
\bibitem {ref37}   L. Gou, J.E. McClintock, J. Liu, et al., Astrophys. J. \textbf{701},1076 (2009)
\bibitem {ref41}   R. Shafee, J.E. McClintock, R. Narayan, et al., Astrophys. J.,\textbf{636}, L113 (2006)
\bibitem {Steineretal2011} J.F. Steiner, R.C. Reis, J.E. McClintock, R. Narayan, et al.,  Mon.Not.R.Astron. Soc. \textbf{416}, 941 (2011) 
\bibitem {ref44}   J.F. Steiner,  J.E. McClintock, J.A. Orosz, et al., Astrophys. J. \textbf{793}, L29 (2014)
\bibitem {ref46}   L. Gou, J.E. McClintock, J.F. Steiner, et al., Astrophys. J. \textbf{718}, L122 (2010)
\bibitem {ref45}   J.F. Steiner,  J.E. McClintock, M.J. Reid, Astrophys. J. \textbf{745}, L7 (2012a)
\bibitem {ref34}   A.C. Fabian, D.R. Wilkins, J.M. Miller, et al., Mon.Not.R.Astron. Soc. \textbf{424}, 217 (2012) 
\bibitem {ref36}   J.M. Miller, M.L. Parker, F. Fuerst,et al., Astrophys. J. \textbf{775}, L45 (2013)
\bibitem {ref38}   J.F. Steiner, R.C. Reis, A.C. Fabian, et al., Mon.Not.R.Astron. Soc. \textbf{427}, 2552 (2012)
\bibitem {Miller2009} J.M. Miller, C.S. Reynolds, A.C. Fabian, G. Miniutti, L.C. Gallo, Astrophys. J.  \textbf{697}, 900 (2009)

\bibitem {ref54}   D. Barret,S. Vaughan, Astrophys. J. \textbf{746},131 (2012)

\bibitem {ref50}   M.A. Abramowicz, W. {Kluz{\'n}iak} , Astron. Astrophys. \textbf{374}, L19  (2001)

\bibitem {Wijnandsetal1999} Wijnands R., Homan J., van der Klis M., 1999, Astrophys. J. \textbf{526}, L33 (1999)

\bibitem {remillardetal2002}   R.A. Remillard, G.J. Sobczak, M.P. Muno, J.E. McClintock, Astrophys. J. \textbf{564}, 962 (2002)
  
\bibitem {ref51}   P. Casella, T. Belloni, L. Stella, Astrophys. J. \textbf{629}, 403 (2005)

\bibitem {ref52}   A. Ingram, C. Done, P.C. Fragile, Mon.Not.R.Astron. Soc. \textbf{397}, L101 (2009)

\bibitem {ref53}   A. Ingram, C. Done, Mon.Not.R.Astron. Soc. \textbf{415}, 2323 (2012)

\bibitem {ref55}   R.P. Fender, J. Homan, T.M. Belloni, Mon.Not.R.Astron. Soc. \textbf{396}, 1370 (2009)

\bibitem {ref56}   S.E. Motta, P. Casella, M. Henze, et al., Mon.Not.R.Astron. Soc. \textbf{447}, 2059 (2015)

\bibitem {ref63}   P.C. Fragile, O.M. Blaes, P. Anninos, J.D. Salmonson, Astrophys. J., \textbf{668}, 417 (2007)

\bibitem {ref57}   P.C. Fragile, O.M. Blaes, Astrophys. J. \textbf{687}, 757 (2008)

\bibitem {Liskaetal2017} M. Liska, C. Hesp, A. Tchekhovskoy, et al., Mon.Not.R.Astron. Soc. in press (arXiv:1707.06619)

\bibitem {ref58}   J. van den Eijnden, A. Ingram, P. Uttley, et al., Mon.Not.R.Astron. Soc. \textbf{464}, 2643 (2017)

\bibitem {IngramDone2012FeK} A. Ingram, C. Done, Mon.Not.R.Astron.Soc. \textbf{427}, 934 (2012)

\bibitem {ref61}   A. Ingram, M. van der Klis, M. Middleton, et al., Mon.Not.R.Astron.Soc.,\textbf{461}, 1967 (2016)

\bibitem {Ingrametal2017} A. Ingram, M. van der Klis, M. Middleton, D. Altamirano, P. Uttley, Mon.Not.R.Astron. Soc. \textbf{464}, 2979 (2017)

\bibitem {ref64}   A. Ingram, M. van der Klis, Mon.Not.R.Astron. Soc. \textbf{446}, 3516 (2015)

\bibitem {Antiaetal2017} H.M. Antia, J.S. Yadav, P.C. Agrawal, J. Verdhan Chauhan, et al., 
Astrophys. J. Suppl. \textbf{231}, 10 (2017)

\bibitem {Gendreauetal2016} K.C. Gendreau, Z. Arzoumanian, P.W. Adkins, C.L. Albert, et al., SPIE \textbf{9905}, 99051H (2016)


\bibitem {ref65}   A. Ingram, T.J. Maccarone, J. Poutanen, H. Krawczynski, Astrophys. J., \textbf{807}, 53 (2015)

\bibitem {ref66}   E.H. Morgan, R.A. Remillard, J. Greiner, Astrophys. J. \textbf{482}, 993 (1997)

\bibitem {vandenEijnden2016}   J. van den Eijnden, A. Ingram, P. Uttley, Mon.Not.R.Astron. Soc. \textbf{458}, 3655 (2016)

\bibitem {Schnittman2005} J.D. Schnittman, Astrophys. J. \textbf{621}, 940 (2005)

\bibitem {Beheshtipouretal2016} B. Beheshtipour, J.K. Hoormann, H. Krawczynski, Astrophys. J. \textbf{826}, 203 (2016)

\bibitem {IngramMaccarone2017} A.R. Ingram, T.J. Maccarone, Mon.Not.R.Astron. Soc. \textbf{471}, 4206 (2017) 

\bibitem {dovciaketal04}  Dov{\v c}iak M., Bianchi S., Guainazzi M., et al., 2004, Mon.Not.R.Astron. Soc. 350, 745


\bibitem {ref67}   T. J. Turner, R. F. Mushotzky, T. Yaqoob, et al., Astrophys. J. \textbf{574}, L123 (2002)

\bibitem {ref68}   K. Iwasawa, G. Miniutti, A.C. Fabian, Mon.Not.R.Astron. Soc. \textbf{355}, 1073 (2004)

\bibitem {ref69}   B. de Marco, K. Iwasawa, M. Cappi, et al., Astron. Astrophys. \textbf{507}, 159 (2009)

\bibitem {karssenetal17}  G. D. Karssen, M.  Bursa, A. Eckart, et al., Mon.Not.R.Astron. Soc. \textbf{472}, 4422 (2017) 

\bibitem {ref21}   M. Dov{\v c}iak, S. Bianchi, M. Guainazzi, V. Karas, G. Matt, Mon.Not.R.Astron. Soc. \textbf{350}, 745 (2004)

\bibitem {ref70}   B.M. Peterson, M.C. Bentz, New Astron. \textbf{50}, 796 (2006)

\bibitem {ref71}   P. Uttley, E.M. Cackett, A.C. Fabian, E. Kara, D.R. Wilkins, Astron. Astrophys. R. \textbf{22}, 72 (2014)

\bibitem {ref20}   L. Stella, Nature \textbf{344}, 747 (1990)

\bibitem{Fabianetal2009} A.C. Fabian, A. Zoghbi, R.R. Ross, et al.  Nature \textbf{459}, 540 (2009)  
  
\bibitem {Zoghbietal2012} A. Zoghbi, A.C. Fabian, C.S. Reynolds, E.M. Cackett, Mon.Not.R.Astron. Soc. \textbf{422}, 129 (2012)
  
\bibitem {Karaetal2016} E. Kara, W.N. Alston, A.C. Fabian, E.M. Cackett, P. Uttley, C.S. Reynolds, A. Zoghbi, Mon.Not.R.Astron. Soc. \textbf{462}, 511 (2016)  

\bibitem {Uttleyetal2011} P. Uttley, T. Wilkinson, P. Cassatella, J. Wilms, K. Pottschmidt, M. Hanke, M. B\"{o}ck, Mon.Not.R.Astron. Soc. \textbf{414}, L60 (2011)

\bibitem{DeMarcoetal2015} B. De Marco, G. Ponti, T. Mu\~{n}oz-Darias, K. Nandra, Astrophys. J. \textbf{814}, 50 (2015) 

\bibitem {ref76}   M.A. Nowak, M. Hanke, S.N. Trowbridge, et al., Astrophys. J. \textbf{728},13 (2011)

\bibitem {ref77}    P. Casella, T.J. Maccarone, K. O'Brien, et al., Mon.Not.R.Astron. Soc. \textbf{404}, L21 (2010)

\bibitem {poutanen&veledina14} J. Poutanen, A. Veledina, Space Sci. Rev. \textbf{183}, 61 (2014)

\bibitem {veledinaetal13} A. Veledina, J. Poutanen, I. Vurum, Mon.Not.R.Astron. Soc. \textbf{430}, 3196 (2013)


\bibitem {ref78}   A.C. Fabian, Annual Rev. Astron. Astrophys. \textbf{50}, 455 (2012)

\bibitem {goosmann&matt11} R.W. Goosmann, G. Matt, Mon.Not.R.Astron. Soc. \textbf{415}, 3119 (2011) 

\bibitem {ref84}   S. Bianchi, R.  Maiolino, G. Risaliti, Advances in Astronomy 2012, 782030 (2012)

\bibitem {ref85}   A. De Rosa, F. Panessa, L. Bassani, et al., Mon.Not.R.Astron. Soc. \textbf{420}, 2087 (2012)

\bibitem {ref86}    G. Torricelli-Ciamponi, P.  Pietrini, G. Risaliti, M. Salvati, Mon.Not.R.Astron. Soc. \textbf{442}, 2116 (2014)

\bibitem {ref87}    F. Marin, V. Karas, D. Kunneriath, F. Muleri, Mon.Not.R.Astron. Soc. \textbf{441}, 3170 (2014)

\bibitem {ref88}   A.G. Markowitz, M. Krumpe, R. Nikutta, Mon.Not.R.Astron. Soc. \textbf{439}, 1403 (2014)

\bibitem {ref89}   F. La Franca, F. Fiore, A. Comastri, Astrophys. J. \textbf{635}, 864 (2005)

\bibitem {marinetal16} F. Marin, R.W. Goosmann, P.O.  Petrucci, Astron. Astrophys. \textbf{591}, A23 (2016) 

\bibitem {DiazTrigoetal2014} M. D\\'iaz Trigo, S. Migliari, J.C.A. Miller-Jones, M. Guainazzi, Astron. Astrophys. \textbf{571}, A76 (2014)

\bibitem {ref92}   B.P. Abbott (LIGO Scientific Collaboration and Virgo Collaboration) Phys. Rev. Lett. \textbf{116}, 131103 (2016)

\bibitem {abbottetal17c} B. P. Abbott, et al., Phys. Rev. Lett. \textbf{119}, 161101 (2017c)

\bibitem {abbottetal17b} B. P. Abbott, et al., Astrophys. J. \textbf{848}, L13 (2017b)

\bibitem {siegel&ciolfi16a} D. Siegel \&  R. Ciolfi R., Astrophys. J., \textbf{819}, 14 (2016a)

\bibitem {siegel&ciolfi16b} D. Siegel \&  R. Ciolfi R., Astrophys. J., \textbf{819}, 15 (2016b)

\bibitem {ref93}   B.F. Schutz, Nature \textbf{323}, 310 (1986)

\bibitem {abbottetal17a} B.P. Abbott, et al., Nature, \textbf{551}, 85 (2017a)

\bibitem {ref94}   E. Berti, et al., Class. Quant. Grav. \textbf{32}, 243001 (2015)

\bibitem {kramer16} M. Kramer, International Journal of Modern Physics D \textbf{25}, 1630029, (2016)

\bibitem {ref95}   C.M. Will, Living Reviews in Relativity 17 (2014)

\bibitem {ref96}   N. Yunes, K. Yagi, F. Pretorius, Phys. Rev. \textbf{D94}, 084002 (2016)

\bibitem {ref97}   Cardoso V., Pani P., 2017, Nat. Astron. \textbf{1}, 586 (2017)

\bibitem {ref99}   V. Cardoso, L. Gualtieri, Class. Quant. Grav. \textbf{33}, 174001 (2016)

\bibitem {ref100}   L.X. Li, E.R. Zimmerman, R. Narayan, J.E. McClintock, Astrophys. J. Suppl. \textbf{157}, 335 (2005)

\bibitem {ref101}   C. Bambi, Rev. Mod. Phys. \textbf{89}, 025001 (2017)

\bibitem {ref102}   T. Johannsen, Class. Quant. Grav. \textbf{33}, 124001 (2016)

\bibitem {ref103}   T.M. Belloni, L. Stella, Space Sci. Rev. \textbf{183}, 43 (2014)

\bibitem {ref104}   A. Maselli, P. Pani, R. Cotesta,  L. Gualtieri, V. Ferrari, L. Stella, Astrophys. J. \textbf{843}, 25 (2017)

\bibitem {ref105}   L. Hui, J.P. Ostriker, S. Tremaine, E. Witten, Phys. Rev. \textbf{D95}, 043541 (2017)

\bibitem {marsh16} 	Marsh, David J. E., Physics Reports,  \textbf{643}, 1-79, (2016).

\bibitem {ref106}   R. Brito, V. Cardoso, P. Pani, Lect. Notes Phys. \textbf{906}, pp.1 (2015) 

\bibitem {ref107}   A. Arvanitaki , S. Dubovsky, Phys. Rev. \textbf{D83}, 044026 (2011)

\bibitem {ref108}   R. Brito, S. Ghosh, E. Barausse, et al., Phys. Rev. Lett. \textbf{119}, 131101 (2017)



%
%

\end{thebibliography}
\end{document}